\documentclass[journal=jpcafh,manuscript=article,11pt]{achemso}
\setkeys{acs}{maxauthors = 0} 
\usepackage[version=3]{mhchem} 
\usepackage{graphicx} 
\usepackage{subcaption}
\usepackage{color}
\usepackage{multirow}
\usepackage{setspace}
\usepackage{natbib}
\usepackage{amsmath}
\usepackage[dvipsnames]{xcolor}
\usepackage{ulem}
\usepackage{tikz}
\usepackage{mathtools}
\usepackage{siunitx}
\usetikzlibrary{decorations.pathreplacing,angles,quotes}

\usepackage{mathtools}
\DeclarePairedDelimiter\bra{\langle}{\rvert}
\DeclarePairedDelimiter\ket{\lvert}{\rangle}
\DeclarePairedDelimiterX\braket[2]{\langle}{\rangle}{#1\,\delimsize\vert\,\mathopen{}#2}

\newcommand{\cm}{cm$^{-1}$}

\newcommand{\ud}{\,\mathrm{d}}

\newcommand{\boltzmann}[1]{e^{- (\beta \hat{H})  #1 }}
\newcommand{\propag}[1]{e^{#1 \mathrm{i}  \hat{H} t / \hbar}}

\DeclareMathOperator{\Tr}{Tr}

\title{Efficient,
direct calculation of reaction rate coefficients based on a partially rearranged
rovibrational Hamiltonian.
A full-dimensional case study of the H$_2$ + D $\rightarrow$ HD + H reaction}

\author{Gábor A. Ecseri}
\affiliation{Institute  of Chemistry, ELTE E\"otv\"os Lor\'and University,
           H-1117 Budapest, P\'azm\'any P\'eter s\'et\'any 1/A, Hungary}
\alsoaffiliation{ELTE Hevesy Gy\"orgy PhD School of Chemistry, H-1117 
           Budapest, P\'azm\'any P\'eter s\'et\'any 1/A, Hungary}
\author{Attila G. Cs\'asz\'ar}
\affiliation{Institute  of Chemistry, ELTE E\"otv\"os Lor\'and University,
           H-1117 Budapest, P\'azm\'any P\'eter s\'et\'any 1/A, Hungary}
\author{Csaba Fábri}
\affiliation{Department of Theoretical Physics, University of Debrecen, P.O. Box 400, H-4002 Debrecen, Hungary}
\email{fabri.csaba@science.unideb.hu} 

\date{\today}

\begin{document}

\singlespacing

\begin{tocentry}
\includegraphics[scale=0.25]{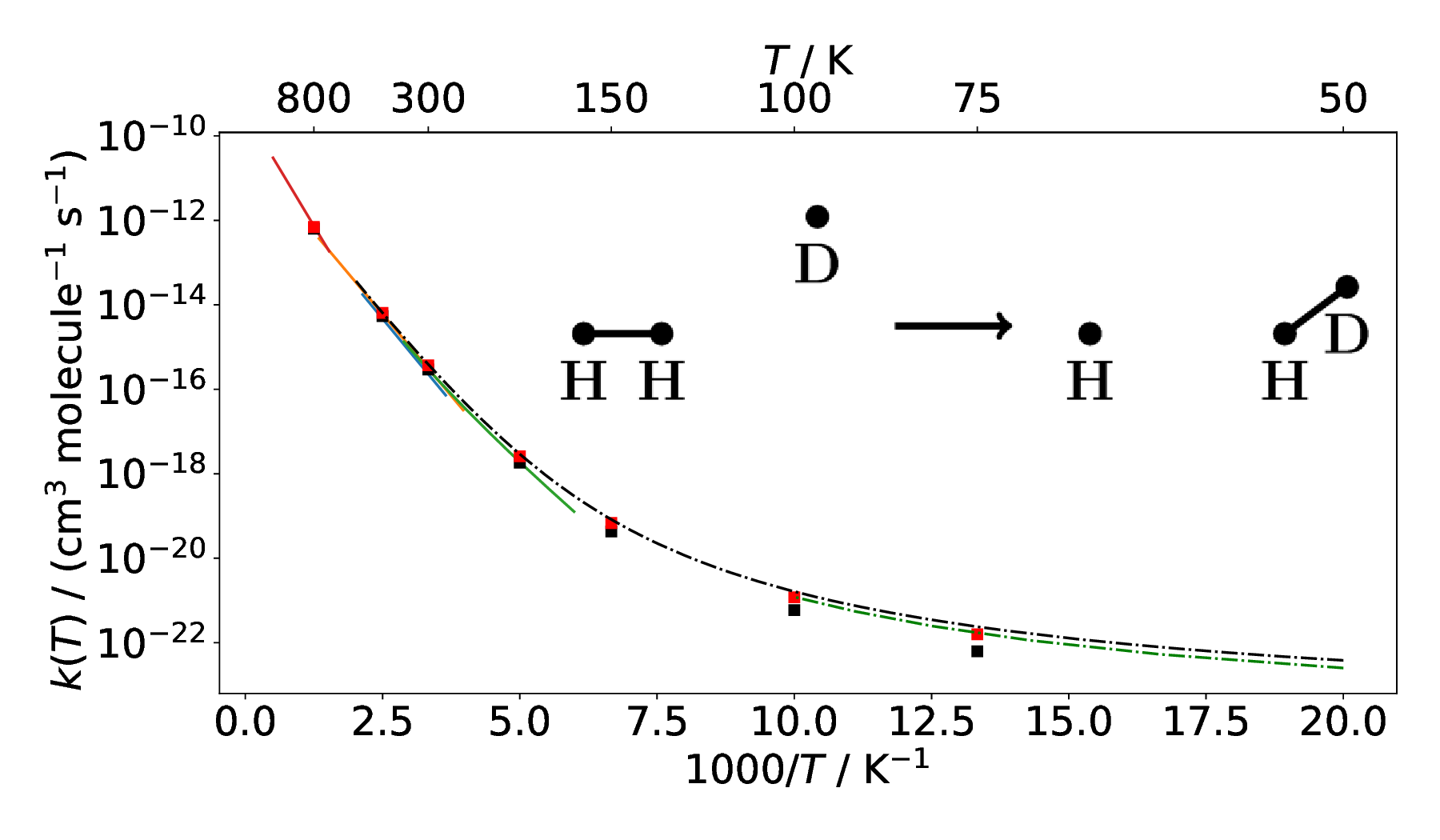}
\end{tocentry}

\begin{abstract}
It is shown that an efficient, `direct', and fully quantum mechanical calculation of thermal reaction rate coefficients requires a new,
partially rearranged form of the numerically-constructed exact kinetic energy part of the rovibrational
Hamiltonian expressed in internal coordinates.
Using this Hamiltonian and an accurate, full-dimensional potential energy surface
characterizing the H$_2$ + H exchange reaction, 
developed by Mielke, Garrett, and Peterson 
(\textit{J. Chem. Phys.} \textbf{2002}, \textit{116}, 4142),
reaction rate coefficients in the 
temperature range of $75-800$~K have been
computed for the 
H$_2$ + D $\rightarrow$ HD + H reaction.
The paper puts particular emphasis on the exact treatment of
overall molecular rotation and on nuclear spin symmetry.
\end{abstract}

\maketitle

\onehalfspacing
\section{Introduction}


During the last century, several models and algorithms have been developed
that allow for the interpretation and computation of thermal
reaction rate coefficients, $k(T)$, of gas-phase reactions.
The most practical approaches include classical and semi-classical
methods \cite{30Eckart,52Marcus,74Miller,75Miller} and methods based on
(variational) transition state 
theory.\cite{31EyPo,35Eyring,35EvPo,38Wigner,59ElHi,83TrHaHy,84TrGa,96TrGaKl}
Since there are several excellent reviews and textbooks covering these approximate
models,\cite{83LaKi,83TrHaHy,84TrGa,96Child,96TrGaKl,97McSi,02ScRa}
we are not considering them any further.
Determination of state-to-state reaction probabilities through the
detailed quantum mechanical study of molecular collision events can also lead to
thermal reaction rate coefficients,\cite{70ChCo,12WeMa}
but for practical purposes, 
this is a somewhat complicated and inefficient approach if no detailed state-to-state information is needed.



It was recognized more than half a century ago \cite{60Yamamoto,74Miller} that
there exists an efficient, `direct' way for the quantum-chemical computation of
thermal reaction rate  coefficients.
This approach is especially useful for chemical reactions
where there are no long-lived resonance states significantly affecting the outcome
of the collision between the reaction partners.

The direct approach to the calculation of $k(T)$ quantum mechanically 
was developed into an efficient algorithm first by Miller \textit{et al}. 
\cite{74Miller,83MiScTr,86Millerb,91SeMi,93Miller,98WaThMi} 
and then by Manthe \textit{et al}.
\cite{95Manthe,98MaMa,98MaMab,00MaMa,07HuMa,11Manthe,12WeHuMa,12WeMa}
There are a few further notable studies based on and advancing this 
formalism.\cite{96ZhLi,97MiCa,01RuCoBa,04MeGr,22KeKaErPe,22DuLaSc}
The principal characteristics of this direct approach, followed in this study, are as follows:
(a) it relates the reaction rate coefficient to the flux through a dividing surface that separates reactants from products;
(b) the central quantity of the formalism is the thermal flux operator, which
has only a relatively small number of non-vanishing eigenvalues; thus,
the corresponding eigenstates can be obtained efficiently and
short-time dynamics is sufficient for the converged determination of the rate;
(c) no separability within the coordinates is required:
there is no ``reaction coordinate'' 
separable from the rest of the coordinates;
(d) no direct information is needed about the asymptotic (reactant and product)
states;
and (e) the dynamical information in a reactive region close to a potential barrier
provides the path to the $k(T)$ coefficients 
without the need for a detailed solution of
the state-to-state reactive scattering problem.\cite{93Miller,98Miller}

There are many possible forms of the exact rovibrational 
Hamiltonian \cite{28Podolsky,76LoGa,07MaCzSuCs,09MaCzCs,11FaMaCs,14Pesonen,15Szalay}
appearing in the equations of the direct method;
in theory, all of them are equivalent.
It is important to emphasize that the
Eckart--Watson \cite{76LoGa,07MaCzSuCs,14Pesonen,15Szalay} form,
which is also exact, cannot be considered in applications where the atoms rearrange.
Forms of the Hamiltonian expressed in internal coordinates should 
almost always be appropriate.
Nevertheless, one must be aware that, in different applications,
different forms of the Hamiltonian may perform differently \cite{09MaCzCs,19AvMa}
as their use may result in different numerical issues.
It has remained unclear what the
best form of the rovibrational Hamiltonian is for 
thermal reaction rate computations.

Despite its formal appeal, practical computations using the formalism developed by
Miller, Manthe, and their co-workers remain somewhat limited.
Probably the most important reasons are as follows:
(a) other ways to compute $k(T)$ are significantly less demanding computationally
when quantum effects are not overwhelmingly important;
(b) general-purpose, black-box-type codes that could be used for
cases involving more than four atoms have not yet been developed;
and (c) the exact inclusion of overall rotational motion seems particularly
cumbersome and expensive.
The present paper addresses the latter two points.

The exact inclusion of overall rotation in direct thermal reaction rate
coefficient calculations can be challenging, even for the H$_3$ system.\cite{92PaLi}
The following are the most typical approximations used for the 
inclusion of rotational excitation in the calculation of reaction rate coefficients:
(a) in the 
$J$-shifting (angular momentum shifting) approach,\cite{91Bowman,97QiBo}
reaction probabilities at non-zero total angular momentum
($J>0$) are related to those at $J=0$ through a shift in the rotational energy
(and consequently the rotational partition function);
and 
(b) during the centrifugal sudden (coupled states)
approximation,\cite{74Pack,74McKo} the rotational-vibrational
coupling term, also called the Coriolis term, is completely neglected.
Testing the limitations of these models (and their generalizations) for various
systems (see, \textit{e.g}., Refs.~\citenum{97QiBo}, \citenum{14PePo},
and \citenum{20SaPoSzCs}) is an interesting topic in itself.

We decided to test the `direct' approach followed during this study
on perhaps the most thoroughly studied chemical reaction, the gas-phase exchange
reaction of H$_2$ + H,\cite{76TrWy,77TrWy,86Schatz}
by applying the method to the isotopic variant \ce{H2 + D -> HD + H}.
There are several reasons for this choice:
(a) studying isotopic variants of the prototypical gas-phase H$_2$ + H
exchange reaction has a
rich history,\cite{30Farkas,30Farkasa,31EyPo,35Farkas,35FaFa,36FaWi,76TrWy,77TrWy}
with particularly strong participation, especially in the early years,
from scientists born in Hungary;
(b) knowledge of a single potential energy surface (PES) appears to be
sufficient, as first pointed out in a ground-breaking paper by London,\cite{29London}
for the accurate computation of its thermal reaction rate coefficients;
(c) there is a significant number of experimental results available, 
\cite{53DaRo, 56BoCaCiMo,59KlMcScSc,65ScRo,66RiScLe,67RoRiQu,67WeHa,68MaPu,72NiMa,73MiRo, 75ApAp,76PrRo,82GlCh,85RoGeIvIl,86DrWo,86MiLe,87ReGaLeSe,90MiFi, 90MiFiBoSu,03MiSuSu}
including ones that showed a curvature in the Arrhenius plot for the
H$_2$ + D reaction at low temperatures;\cite{68MaPu,66RiScLe}
(d) isotopic variants of the H$_2$ + H reaction have been subjected to
computational studies of varying
sophistication,\cite{31EyPo,59Shavitt,59Weston,66RiScLe,80GaTr,85BlTr, 86GaTrSc,86WeLi, 87PaPa,87ScHaTrSu,87TaMaNaOk,87ZhMi,88Schatz,91PaLi,94MiLyTrSc, 96AoBaDiHe,97BaMe,01Kendrick,02SuMe,03Kendrick,05AoBaHe,20XiZhWa,23WaHuWa,24LiHuLu,26FaHuZh}
some indicating that, using less accurate PESs, there are
deviations between the experimental and the computational reaction-rate studies,
both at low (200 K) and high  (1500 K) temperatures;
and (e) at the transition state, both the vibrational and the rotational
wavenumbers are relatively large,
facilitating the convergence of the `direct' modeling approach.

Computing thermal reaction rates requires a PES.
Over the last 60 years, several full-dimensional PESs
have been computed for the three-electron H$_3$ 
system \cite{73Liu,78SiLi,78TrHo,79TrHo,79Varandas,87VaBrMeTr,91BoKeMaPe,96BoKeMaPe, 99WuKuAn,02MiGaPe,25GhRoKuNa}
(for a historical overview of the earlier PESs see 
Refs.~\citenum{77TrWy} and \citenum{02MiGaPe}).
These PESs include, in alphabetical order, the BKMP \cite{91BoKeMaPe},
BKMP2,\cite{96BoKeMaPe} CCI,\cite{02MiGaPe} DMBE,\cite{87VaBrMeTr}
EQMC,\cite{99WuKuAn} LEPS,\cite{79Varandas} and
LSTH \cite{73Liu,78SiLi,78TrHo,79TrHo} variants.
In this paper, results are reported only for the CCI PES of Ref.~\citenum{02MiGaPe}.
Within the standard Born--Oppenheimer \cite{27BoOp} approximation,
these PESs are independent of the masses and nuclear spins.
Corrections due to the diagonal Born--Oppenheimer correction (DBOC)\cite{86HaYaSc,03VaSh}
were also reported,\cite{02MiGaPe} 
they have particular relevance for the lightest molecular systems.
Note that a significant number of dynamical results
(cross sections, reaction rate constants, and resonances) from
statistical, reduced-dimensionality, semiclassical, and full quantal studies 
have been presented using the BKMP, BKMP2, CCI, DMBE, and LSTH
surfaces.\cite{80GaTr,80GaTrGrWa,83WaHa,88Schatz,88ZhMi,89BlZhMlTr,91PaLi, 94MiLyTrSc,94WaBo,97BaMe,98BaAoHe,01AoBaCa, 01Kendrick,03MiPeScGa,05AoBaHe,13SuDeJaCa,17RaGh}

Experimental measurements of the thermal reaction rate coefficients of the 
H$_2$ + D $\rightarrow$ HD~+~H reaction are available over a temperature range
of $167-2112$~K.
\cite{56BoCaCiMo,59KlMcScSc,66RiScLe,67WeHa,72NiMa,73MiRo,75ApAp,76PrRo,82GlCh,84RoGeIvKu,85RoGeIvIl,86DrWo,90MiFi,90MiFiBoSu,03MiSuSu,03MiPeScGa}
Considering the variety of experimental techniques used to produce $k(T)$ values,
most of them agree reasonably well.
Particularly useful experimental data \cite{90MiFi,90MiFiBoSu} have been provided
by the flash photolysis--shock tube technique. 
The experimental data have provided a fertile and stringent testing ground for
the computational studies. 


Finally, 
the hallmarks of the present computational study: 
(a) it takes advantage of a fully numerical approach offered by the general,
black-box type \texttt{GENIUSH} algorithm and code,\cite{09MaCzCs,11FaMaCs}
allowing the straightforward use of
arbitrary internal coordinates, body-fixed frame embeddings, and reduced-dimensional models
for exploring the power of the `direct' method to
compute thermal reaction rate coefficients;
(b) it uses a new form of the exact kinetic energy operator, 
called the partially rearranged form, which is needed to obtain correct
reaction rate coefficients;
(c) it includes the rotational and nuclear spin degrees of freedom in the
numerically exact computations,
which are crucial for thermal reaction rate computations; 
and (d) it yields results that are in good agreement with earlier experimental
and computational studies for the \ce{H2 + D -> HD + H} reaction.

\section{Methodology 
}

\subsection{The `direct' method} 

As foreseen by Yamamoto \cite{60Yamamoto} and elaborated by
Miller \textit{et al}.,\cite{83MiScTr} thermal reaction rate coefficients of
bimolecular reactions at temperature $T$  can be obtained from the standard
flux autocorrelation function $C^T_{\rm f}(t)$ as a time ($t$) integral:
\begin{equation}
   k(T) = \frac{1}{Q_{\rm r}(T)} \int_0^\infty C^T_{\rm f}(t) \ud t ,
   \label{ArtEq:kTint}
\end{equation}
where $Q_{\rm r}(T)$ is the partition function of the reactants (per unit volume).
In a quantum mechanical context, the standard time-dependent flux autocorrelation
function of Eq.~\eqref{ArtEq:kTint} can be expressed as
\begin{equation}
    C^T_{\rm f}(t) = \Tr[\boltzmann{/2} \hat{F} \boltzmann{/2} \propag{} \hat{F} \propag{-} ] ,
\label{ArtEq:CffTrace}
\end{equation}
where $\hat{H}$ is the (rovibrational) Hamiltonian of the system,
$\hat{F}$ is the reactive flux operator,
$\hbar$ is the reduced Planck constant, and $\beta=1/(kT)$, where $k$ is the Boltzmann constant and $T$ is the thermodynamic temperature. 
The reactive flux operator is defined as
$\hat{F} := \mathrm{i}/\hbar [\hat{H}, \theta(s(\textbf{q}))]$
where $\theta(x) = \begin{cases}
    0,& x < 0 \\
    1,& x \ge 0
\end{cases}$ is the Heaviside step function.
The reactant and product configurations are separated by a dividing surface, $s(\textbf{q})=0$. 
The multivariate function $s$(\textbf{q}), dependent on the internal coordinates $\textbf{q}$ chosen for the system,
has negative/positive values for the reactant/product configurations. 

The quantum mechanical trace in Eq. (\ref{ArtEq:CffTrace}) should be evaluated
using a suitable basis; 
in the present work, we utilize a methodology in which the trace is evaluated
in the eigenbasis of the so-called thermal flux operator.\cite{88PaLi,12WeMa}
The thermal (sometimes called Boltzmannized \cite{88PaLi,97ThMi})
flux operator is defined  as
\begin{equation}
\label{eq:thermalflux}
    \hat{F}(\beta) := \boltzmann{/2} \hat{F} \boltzmann{/2}.
\end{equation}
Then, Eq. (\ref{ArtEq:CffTrace}) can be rewritten as
\begin{equation}
    C^T_{\rm f}(t) = \sum_i f_i \bra{\phi_i(t)}\hat{F}\ket{\phi_i(t)},
\label{eq:cft}
\end{equation}
where $\hat{F}(\beta) \ket{\phi_i} = f_i \ket{\phi_i}$ and $\ket{\phi_i(t)} = \propag{-}\ket{\phi_i}$. 
As pointed out before, see, \textit{e.g.}, Ref. \citenum{97ThMi},
$\hat{F}(\beta)$ is a low-rank operator,
and its eigenvectors corresponding to the largest (in absolute value) eigenvalues
provide a compact basis to evaluate the operator trace in Eq. \eqref{ArtEq:CffTrace}
and obtain a converged value based on just a relatively small number of
eigenstates.

Once the thermal flux eigenpairs are obtained at a single temperature, $T$,
the standard flux autocorrelation function at $T/2$ can also be computed
straightforwardly as \cite{98WaThMi}
\begin{equation}
C_{\rm f}^{T/2}(t) = \sum_{ij} f_i f_j |\langle\phi_j |\phi_i(t)\rangle|^2.
\label{ArtEq:halftemp}
\end{equation}
Since
$\langle\phi_j | \propag{-}\phi_i\rangle = \langle (e^{-\textrm{i} \hat{H} t/(2\hbar)} \phi_j^*)^* |e^{-\textrm{i} \hat{H} t/(2\hbar)} \phi_i \rangle$,
by propagating the thermal flux eigenvectors up to $t/2$,
$C_{\rm f}^{T/2}$ can be obtained at time $t$ \cite{98MaMa}. 
Note that in the literature, several other, more complicated techniques have been
described that can be used to compute reaction rate coefficients
at several different temperatures from a
single diagonalization of the thermal flux at a chosen reference 
temperature.\cite{95Manthe,98MaMa,04MeGr}

Since the thermal flux operator commutes with the overall angular momentum operator
$\hat{J}$, the thermal flux can be diagonalized for each rotational quantum number
$J$ separately; that is, $\hat{F}^J(\beta) \ket{\phi_i^J} = f_i^J \ket{\phi_i^J}$.
Thus, separate flux autocorrelation functions $C_{\rm f}^{T,\,J}(t)$ 
and reaction rate coefficients $k^J(T)$ can be computed. 
The full thermal reaction rate coefficient is a simple sum of these:
\begin{equation}
    k(T) = \sum_{J} (2J+1)k^J(T).
\label{eq:ktrot}
\end{equation}

If indistinguishable atoms are present in the reactive system,
we must take the Pauli exclusion principle into consideration.
Thus, in the computation of the reaction rate coefficient,
the basis used to evaluate the trace of Eq. (\ref{ArtEq:CffTrace})
must be symmetric or antisymmetric with respect to the exchange of identical bosons or fermions, respectively.
In the present application, the basis consists of the product of
thermal flux eigenstates and nuclear spin states.
Earlier, nuclear spin states were not mentioned,
as the operators within the trace of Eq. (\ref{ArtEq:CffTrace}) do not act
on the spin coordinates. 
However, it is now necessary to include the nuclear spin states of the system,
since they become relevant when the Pauli exclusion principle is taken into account.

The effects caused by the involvement of identical atoms in a reaction,
that is, the consideration of nuclear spin states, should be considered for all
collisions and binary reactions.\cite{63Schlag,77Quack}
For example, in the case of H$_2$ in the \ce{H2 + D -> HD + H} reaction,
there are three symmetric (\textit{ortho}, $I=1$,
where $I$ denotes the nuclear spin quantum number) nuclear spin state functions:
$(1/\sqrt{2})(\alpha(1)\beta(2)+\alpha(2)\beta(1))$, $\alpha(1) \alpha(2)$,
and $\beta(1) \beta(2)$ and
there is one antisymmetric (\textit{para}, $I=0$)
nuclear spin state function, $(1/\sqrt{2})(\alpha(1)\beta(2)-\alpha(2)\beta(1))$. 
$\beta(1)$/$\alpha(1)$ denotes that H$^{(1)}$ is in $\beta$/$\alpha$
nuclear spin states, respectively, and similarly for H$^{(2)}$.
The thermal flux eigenstates have a well defined symmetry property only if the thermal flux operator 
commutes with all the permutation operators of the equivalent atoms, $\hat{P}_{ij}$. 
The commutator of $\hat{P}_{ij}$ and the thermal flux can be expressed as
\begin{equation}
    [\hat{P}_{ij},\hat{F}(\beta)] = \boltzmann{/2} [\hat{P}_{ij},\hat{F}] \boltzmann{/2} = 
    \frac{\textrm{i}}{\hbar} \boltzmann{/2} [\hat{H},[\hat{P}_{ij}, \theta(s(\textbf{q}))]] \boltzmann{/2}.
\end{equation}
Here, we have exploited the fact that $[\hat{P}_{ij},\hat{H}] = 0$.
Therefore, if the dividing surface is chosen so that $[\hat{P}_{ij},
\theta(s(\textbf{q}))]=0$, $\hat{P}_{ij}$ commutes with the thermal flux.

In the case of the \ce{H2 + D -> HD + H} reaction,
it is possible to guarantee that $[\hat{P}_{12}, \theta(s(\textbf{q}))]=0$.
Therefore, thermal flux eigenstates are either symmetric or antisymmetric under
the exchange of H$^{(1)}$ and H$^{(2)}$ ($\hat{P}_{12}$).
To enforce the Pauli exclusion principle, symmetric/antisymmetric thermal flux
eigenstates must be paired with antisymmetric/symmetric (\textit{para}/\textit{ortho})
nuclear spin states of the two protons.
This is the basis used in this study to evaluate the trace of
Eq. \eqref{ArtEq:CffTrace}, which can be carried out separately using only the
\textit{ortho} or only the \textit{para} states, 
resulting in the $C_{\rm f}^{\textit{ortho}}(t)$ or $C_{\rm f}^{\textit{para}}(t)$
flux autocorrelation functions, respectively,
and finally in the $k^{\textit{ortho}}(T)$ or $k^{\textit{para}}(T)$ 
reaction rate coefficients.
Since the operators within the trace are independent of the nuclear spin, 
the three $k^{\textit{ortho}}(T)$ contributions are equal, yielding a
spin statistical factor of $g=3$.
Generally, the full thermal reaction rate coefficient can be written as
the sum of the contributions from the different nuclear spin states,
\begin{equation}
    k(T) = \sum_i g_i k^i(T),
\end{equation}
and in the specific case of the \ce{H2 + D -> HD + H} reaction,
\begin{equation}
   k(T) = 3 k^\textit{ortho}(T) + k^\textit{para}(T).
\label{eq:ktspin}
\end{equation}

\subsection{GENIUSH}
The set of codes \texttt{GENIUSH},\cite{09MaCzCs,11FaMaCs,12CsFaSzMa,14FaMaCs, 17FaQuCs,17PaSzCs,19SiSzCs,23SiFaCs,23SiScBrFa}
upon which the present study is based, is a general, black-box type,
completely numerical quantum-chemical code designed \cite{09MaCzCs,11FaMaCs}
to solve the time-independent nuclear
Schr\"odinger equation for rovibrational energy levels and eigenstates
in a (quasi-)variational fashion.
\texttt{GENIUSH} uses the exact and general kinetic energy operator 
constructed numerically;
therefore, no analytical form of the kinetic energy is needed.
\texttt{GENIUSH} can accommodate any set of internal coordinates,
any embedding of the 
molecular (body-fixed) axes, and can solve arbitrary,
reduced- or full-dimensional molecular models.

In \texttt{GENIUSH}, 
the matrix representation of the rovibrational Hamiltonian is set up
in a multidimensional direct-product discrete-variable
representation (DVR) \cite{00LiCa} basis.
Since every coordinate-dependent operator $f(\textbf{q})$ has
a diagonal matrix representation in DVR and the diagonal entries are simply
the values of $f(\textbf{q})$ at the  DVR points,
the numerical treatment of the potential energy surface and the dividing surface
is greatly simplified.
Due to these characteristics,
\texttt{GENIUSH} provides an ideal starting point for the
development of a general code aimed at the computation of reaction rate
coefficients.

The most time-consuming step of a \texttt{GENIUSH} computation is the 
determination of a given set of energy levels and eigenstates of the
rovibrational Hamiltonian.
In \texttt{GENIUSH}, the sparse DVR Hamiltonian matrix is diagonalized by
the iterative Lanczos algorithm,\cite{50Lanczos,85CuWi}
which requires the evaluation of matrix-vector products. 
While the `direct' determination of reaction rate coefficients does not require
the eigenpairs of the rovibrational Hamiltonian, 
it does need those of the thermal flux, and the sparse DVR representation of the
Hamiltonian can be employed for the representation of the thermal flux.

\subsection{The partially rearranged form of the kinetic energy operator}
It is the so-called Podolsky form \cite{28Podolsky} of the exact kinetic energy
operator that has usually been used in
large-scale computations of bound rovibrational quantum states performed with
\texttt{GENIUSH}.
To obtain correct reaction rate coefficients, it has become necessary to
rearrange the Podolsky form of the rovibrational kinetic energy operator.
The numerical issues experienced with the Podolsky form of the
Hamiltonian and the solutions to them are outlined in the Appendix.
The new form corresponds to a rearrangement of the vibrational kinetic energy term
$\hat{T}^\textrm{vib}$ in the Podolsky form, while
the rotational and rovibrational coupling kinetic energy terms
($\hat{T}^\textrm{rot}$ and $\hat{T}^\textrm{rotvib}$, see Eqs. (19) and (20)
of Ref.~\citenum{11FaMaCs}, respectively) remain unchanged.

The Podolsky form of the vibrational kinetic energy operator $\hat{T}^\textrm{vib}$
in a set of internal coordinates $q_k$ is
\begin{equation}
    \hat{T}^\textrm{vib} = \frac{1}{2} \sum_{k,l=1}^{D} \tilde{g}^{-1/4} \hat{p}_k^\dagger G_{kl} \tilde{g}^{1/2} \hat{p}_l \tilde{g}^{-1/4} ,
\label{eq:tvibpod}
\end{equation}
where the momentum operator is defined as $\hat{p}_k := - \textrm{i} \hbar \frac{\partial}{\partial q_k}$
(see Ref. \citenum{11FaMaCs} for the definition of the quantities $G_{kl}$ and $\tilde{g}$).
One can divide the double sum in Eq. \eqref{eq:tvibpod} into two terms,
one with $k=l$ and one where $k \ne l$:
\begin{gather}
    \hat{T}^\textrm{vib} = - \frac{\hbar^2}{2}  \tilde{g}^{-1/4} \bigg[ \sum_{k = 1}^{D} \frac{\partial}{\partial q_k} G_{kk} \tilde{g}^{1/2} \frac{\partial}{\partial q_k} + \sum_{k \neq l}^{D}\frac{\partial}{\partial q_k} G_{kl} \tilde{g}^{1/2} \frac{\partial}{\partial q_l} \bigg] \tilde{g}^{-1/4} \nonumber \\
    = - \frac{\hbar^2}{2}  \tilde{g}^{-1/4} \bigg[ \sum_{k = 1}^{D} \left( \frac{\partial (G_{kk} \tilde{g}^{1/2})}{\partial q_k}  \frac{\partial}{\partial q_k} + G_{kk} \tilde{g}^{1/2} \frac{\partial^2}{\partial q_k^2} \right) + \sum_{k \neq l}^{D}\frac{\partial}{\partial q_k} G_{kl} \tilde{g}^{1/2} \frac{\partial}{\partial q_l} \bigg] \tilde{g}^{-1/4}.
\label{ArtEq:prearr_deriv}
\end{gather}
During the second step in Eq. \eqref{ArtEq:prearr_deriv}, 
the first $\frac{\partial}{\partial q_k}$ derivative is evaluated using the
product rule, acting on both $G_{kk} \tilde{g}^{1/2}$ and the subsequent derivatives.
The second line of Eq. \eqref{ArtEq:prearr_deriv} is referred to as the
partially rearranged form used to compute reaction rate coefficients in this study.
Considerations regarding the hermiticity of the matrix representation of the
partially rearranged form are described in the Appendix.
In summary, the partially rearranged form of the kinetic energy operator is
similar to the well-known Podolsky form;
only the diagonal terms ($k=l$) are rearranged in $\hat{T}^\textrm{vib}$.

\subsection{The thermal flux operator}
The thermal flux operator $\hat{F}(\beta)$, see Eq. \eqref{eq:thermalflux},
is purely imaginary and Hermitian;
thus, it has eigenvalue pairs of opposite signs and equal absolute values,
along with eigenfunctions that are complex conjugates of one another.
As detailed in Ref.~\citenum{02Manthe}, for example,
each pair can be considered to correspond to a rovibrational state of the
activated complex.
The numerical treatment and diagonalization of the thermal flux operator,
a step required in the algorithm used,
can be accelerated by tricks similar to those employed during the development of the
\texttt{GENIUSH} code.\cite{09MaCzCs,11FaMaCs}

If an iterative eigensolver is used to obtain the eigenvalues and eigenvectors of the thermal flux, 
matrix-vector products need to be evaluated.
The product of the matrix representation of the thermal flux, $\mathbf{F}(\beta)$,
with a vector, $\mathbf{v}$, is carried out in three steps,
\begin{equation}
    \mathbf{F}(\beta)  \mathbf{v} = e^{-(\beta \mathbf{H})/2 } \left(\mathbf{F}  \left(e^{-(\beta \mathbf{H})/2} \mathbf{v} \right)  \right ),
\end{equation}
where $\mathbf{H}$ and $\mathbf{F}$ are the DVR matrix representations of the 
Hamiltonian and the flux operator, respectively.
The first and last steps are multiplications by the matrix representation
of the Boltzmann operator,
which can be regarded as time propagation in imaginary time.
Therefore, common time propagation techniques can be used. 
In this work, we use the so-called Chebyshev expansion of the
propagator \cite{84TaKo,91LeBiCeFe,94Kosloff}. 
In principle, other propagation methods could also be used,
such as the split-operator technique. 
However, we employed the Chebyshev propagation, as its implementation in 
\texttt{GENIUSH} is straightforward.
Since the Chebyshev propagation requires only the evaluation of
matrix-vector products $\mathbf{H} \mathbf{v}$,
it is not necessary to explicitly construct and store the Hamiltonian matrix,
which can become prohibitively large.
The second step is the multiplication by the matrix of the reactive flux. It is performed according to
\begin{equation}
    \mathbf{F} \mathbf{v} = \frac{\rm i}{\hbar} (\mathbf{H}\boldsymbol{\theta}\mathbf{v} - \boldsymbol{\theta}\mathbf{H}\mathbf{v}),
\end{equation}
where $\boldsymbol{\theta}$ is the diagonal matrix of $\theta(s(\textbf{q}))$
evaluated in the multidimensional direct-product DVR basis used to represent the Hamiltonian.  
In total, one action of the thermal flux needs $2N+2$ actions of the Hamiltonian,
where $N$ is the number of terms needed in the Chebyshev expansion,
typically not exceeding a few hundred.

The ARPACK package \cite{ARPACK} was used to perform the diagonalization of the
thermal flux matrix.
ARPACK is a Fortran library that uses the Arnoldi method to iteratively
diagonalize complex matrices. 
Thermal flux eigenstates corresponding to the dominant (largest in absolute value)
eigenvalues of $\mathbf{F}(\beta)$ were then propagated in time using the Chebyshev
expansion method, and the flux autocorrelation function was evaluated by
Eq.~\eqref{eq:cft} for an appropriately chosen time interval.

\section{Computational details}

\begin{figure}[b!]
\centering
\includegraphics[]{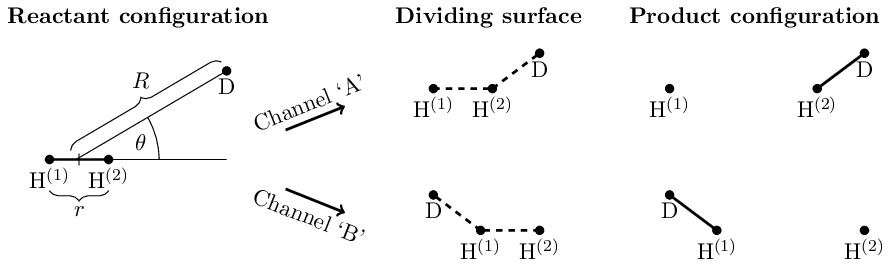}
\caption{Schematic representation of the dividing surface as well as
the reactant and product configurations of the \ce{H2 + D -> HD + H} reaction.
The definition of the Jacobi coordinates $r$, $R$, and $\theta$ is also displayed.
Channel `A' describes the reaction in which HD is formed with H$^{(2)}$,
while the reaction resulting in H$^{(1)}$D corresponds to channel `B'.
The two channels are physically indistinguishable.
}
\label{ArtFig:3D_tikz}
\end{figure}

This study utilizes Jacobi (also called scattering) coordinates,\cite{1842Jacobi}
shown in Fig.~\ref{ArtFig:3D_tikz}.
They are defined as follows:
$r$ is the distance between the two hydrogen atoms,
$R$ is the distance between the deuterium and the center of mass (COM) of 
H$^{(1)}$ and H$^{(2)}$, while
$\theta$ is the angle formed by the three points $\mathbf{r}_\mathrm{2}$,
$\mathbf{r}_\mathrm{COM(1,2)}$, and $\mathbf{r}_\mathrm{D}$, 
where $\mathbf{r}_\mathrm{2}$ and $\mathbf{r}_\mathrm{D}$ 
are the positions of H$^{(2)}$ and D, respectively, and
$\mathbf{r}_\mathrm{COM(1,2)}$ is the position of the 
center of mass of H$^{(1)}$ and H$^{(2)}$.
Figure~\ref{ArtFig:3D_tikz} also shows
the schematic representation of the two reaction channels.
The body-fixed frame is defined by directing the first Jacobi vector ($\mathbf{r}_2-\mathbf{r}_1$) along the positive $z$ 
axis and placing the second Jacobi vector ($\mathbf{r}_\mathrm{D}-\mathbf{r}_\mathrm{COM(1,2)}$) in the $xz$ plane (with $x\ge0$).
Under the permutation of the two equivalent hydrogen atoms,
the Jacobi coordinates transform as
$(r,R,\theta) \xrightarrow{\hat{P}_{12}} (r,R,\pi-\theta)$ and a $180 ^{\circ}$ rotation about the body-fixed $x$ axis
is needed to restore the definition of the body-fixed frame.

The dividing surface of the \ce{H2 + D -> HD + H} reaction is
defined such that either the distance between H$^{(2)}$ and D, or H$^{(1)}$ and D,
is equal to the distance between the two hydrogen atoms. 
All configurations where the H$^{(2)}$-D or the H$^{(1)}$-D distance is smaller
than the H$^{(2)}$-H$^{(1)}$ distance are considered products. 
All the other configurations are classified as reactants.
Most importantly, this dividing surface is 
invariant to the permutation of the two hydrogen atoms ($\hat{P}_{12}$); thus, eigenfunctions of the thermal flux operator are either symmetric or antisymmetric with respect to $\hat{P}_{12}$.
The symmetry properties of the thermal flux eigenfunctions under $\hat{P}_{12}$ were determined numerically by examining the eigenvectors of the thermal flux matrix $\mathbf{F}(\beta)$.

\begin{table}[t!]
    \centering
    \begin{tabular}{ccccccc}\hline \hline
        & & & \multicolumn{3}{c}{grid size} &  \\
        \cline{4-6}
        $T$/K &  $J_{\textrm{max}}$ & int. time limit/a.u. & $r$ & $R$ &  $\theta$ & upper $r/R$ limit/\AA\\ \hline
        100 & 8 & 1900 & 30 & 30 & 30 & 3.2\\
        150/75 & 8 & 1000/2500 & 36 & 36 & 30 & 3.6\\
        200/100 & 9 & 1000/1900 & 30 & 30 & 30 & 3.2\\
        300 & 12 & 1000 & 20 & 20 & 30 & 2.5\\
        800/400 & 20 (12) & 800 & 30 & 30 & 30 & 2.5\\ \hline \hline
    \end{tabular}
    \caption{Selected parameters of the reaction-rate computations.
    In the case of $T=800/400$ K, computations were performed for
    $J=0,...,12,20$ and values of $k^J(T)$ were then obtained \textit{via}
    interpolation between $J=12$ and $20$.
    ``int. time limit'' is the upper integration time limit used to integrate the flux
    autocorrelation function $C^T_{\rm f}(t)$.}
    \label{ArtTab:compparam}
\end{table}

For the computation of thermal reaction rate coefficients, we have used the
CCI PES developed by Mielke, Garrett, and Peterson.\cite{02MiGaPe} 
A potential cutoff of 20\,000 \cm{} was applied to regularize the highest
eigenvalue of the Hamiltonian, as this makes the 
Chebyshev expansion \cite{84TaKo,91LeBiCeFe,94Kosloff} employed in this study
more efficient.
In all computations, atomic masses 
$m_\textrm{H} = 1.007\,825~\textrm{u}$ and 
$m_\textrm{D} = 2.014\,102~\textrm{u}$
have been employed.

Sine-DVR bases\cite{92CoMi}
were used along the $r$ and $R$ Jacobi coordinates (Fig.~\ref{ArtFig:3D_tikz}).
The number of basis functions and the upper limit of the grid were varied
for the different computations, 
while the lower limits of the grid were set to 0.10 \AA{} and 0.75 \AA{} for the
$r$ and $R$ coordinates, respectively.
Legendre-DVR basis functions were used along the $\theta$ coordinate,
with DVR points distributed in the range $[0^{\circ},180^{\circ}]$.
With these grid settings, both reaction channels shown in Fig.~\ref{ArtFig:3D_tikz}
are incorporated into our computations. 
The grid sizes and the upper limits of the grid along the $r$ and $R$ coordinates
for the different computations can be found in Table~\ref{ArtTab:compparam}.
For the rotational degrees of freedom,
Wang combinations of symmetric top eigenfunctions were employed.\cite{11FaMaCs}
Table~\ref{ArtTab:compparam} also contains the upper time integration limit
used to integrate the flux autocorrelation function.
With these settings, we have found that the flux autocorrelation function
remains zero for a sufficiently long time before unphysical reflections from
the edges of the grid set in.
Therefore, choosing an appropriate upper time integration limit posed
no difficulty, and no complex absorbing potential (CAP)\cite{11Moiseyev}
was applied during the present work.

The determination of thermal reaction rate coefficients requires
accurate reactant partition functions, $Q_{\rm r}(T)$, which can be readily evaluated by the direct summation technique.
Accurate ideal-gas partition functions are available for a large number of
molecular systems; see, \textit{e.g.}, Refs.~\citenum{17GaRoLoGo} and
\citenum{25TuZsPaNa}.
For the H$_2$ molecule, Refs.~\citenum{17GaRoLoGo} and \citenum{16PoJo}
provide particularly accurate thermochemical data.
The partition function of the reactants in the  \ce{H2 + D -> HD + H} reaction can be expressed as
\begin{equation}
    Q_{\rm r}(T) = Q_{\rm r}^{\rm transl}(T)Q_{\rm r}^{\rm int}(T) =
    \left(\frac{\mu}{2\pi \hbar^2 \beta}\right)^{3/2}Q_{\rm r}^{\rm int}(T) 
\end{equation}
where 
$\mu$ is the reduced mass of H$_2$ and D.
For this study, the internal partition function $Q_{\rm r}^{\rm int}(T)$ of H$_2$
was taken from Ref.~\citenum{17GaRoLoGo}.

The computations of the thermal rate coefficients were performed at
$T=100$, 150, 200, 300, and 800~K for all
$J$ values up to $J_{\rm max}$ (see Table~\ref{ArtTab:compparam} for which
$J_{\rm max}$ value was chosen for a given temperature).
The effects of rotation and nuclear spin were taken into account
using Eq. \eqref{eq:ktrot} and Eq. \eqref{eq:ktspin}, respectively. 
Thermal reaction rate coefficients were also obtained at
$T=75$, 100, and 400~K \textit{via} the use of Eq.~\eqref{ArtEq:halftemp}. 
Thus, the reaction rate coefficient at 100 K was computed by two different methods,
enabling a check of the performance of Eq. \eqref{ArtEq:halftemp}.

Despite the fact that optimization of the code has not been our major focus, for $J = 0$, the calculation time is trivial for the \ce{H2 + D -> HD + H} reaction; it takes only a few minutes of CPU time.
The two most expensive parts of the reaction rate calculations are the Arnoldi diagonalization and the time evolution; the diagonalization typically takes about $10-20 ~ \%$ of the total CPU time (slightly more for the lowest $J$ values).
The CPU requirement of the benchmark-type $J > 0$ test calculations increases rapidly with the $J$ values, CPU times typically range from one hour to one week (using $32$ threads with OpenMP parallelization).

\section{Results and discussion}

\begin{table}[t!]
    \centering
\resizebox{1.0\columnwidth}{!}{
    \begin{tabular}{cccccc|ccccc}\hline \hline
            & \multicolumn{5}{c|}{Without barrier correction} & \multicolumn{5}{c}{With barrier correction} \\ \cline{2-6}\cline{7-11}
        $T$/K &  $k^\textit{ortho}(T)$ & $k^\textit{para}(T)$ & $k^\textit{para}/k^\textit{ortho}$ & $k^{\rm full}$ & $k^{\rm extrapolated}$ &  $k^\textit{ortho}(T)$ & $k^\textit{para}(T)$ & $k^\textit{para}/k^\textit{ortho}$ & $k^{\rm full}$ & $k^{\rm extrapolated}$ \\ \hline
75 & \num{3.4986e-23} & \num{5.3614e-23} & \num{1.53} & \num{1.5857e-22} & \num{1.5886e-22} & \num{1.3768e-23} & \num{2.1098e-23} & \num{1.53} & \num{6.2400e-23} & \num{6.2515e-23}\\
100 & \num{2.8802e-22} & \num{3.2810e-22} & \num{1.14} & \num{1.1922e-21} & \num{1.1984e-21} & \num{1.4310e-22} & \num{1.6301e-22} & \num{1.14} & \num{5.9231e-22} & \num{5.9540e-22}\\
$100=200/2$ & \num{2.8924e-22} & \num{3.2995e-22} & \num{1.14} & \num{1.1977e-21} & \num{1.1996e-21} & \num{1.4371e-22} & \num{1.6393e-22} & \num{1.14} & \num{5.9505e-22} & \num{5.9600e-22}\\
150 & \num{1.7061e-20} & \num{1.7104e-20} & \num{1.00} & \num{6.8288e-20} & \num{6.9362e-20} & \num{1.0703e-20} & \num{1.0730e-20} & \num{1.00} & \num{4.2837e-20} & \num{4.3511e-20}\\
200 & \num{6.4825e-19} & \num{6.4836e-19} & \num{1.00} & \num{2.5931e-18} & \num{2.6292e-18} & \num{4.5693e-19} & \num{4.5701e-19} & \num{1.00} & \num{1.8278e-18} & \num{1.8533e-18}\\
300 & \num{9.2821e-17} & \num{9.2833e-17} & \num{1.00} & \num{3.7130e-16} & \num{3.7326e-16} & \num{7.3517e-17} & \num{7.3527e-17} & \num{1.00} & \num{2.9408e-16} & \num{2.9563e-16}\\
400 & \num{1.6119e-15} & \num{1.6123e-15} & \num{1.00} & \num{6.4480e-15} & \num{6.4482e-15} & \num{1.3533e-15} & \num{1.3536e-15} & \num{1.00} & \num{5.4135e-15} & \num{5.4136e-15}\\
800 & \num{1.7385e-13} & \num{1.7387e-13} & \num{1.00} & \num{6.9541e-13} & \num{6.9846e-13} & \num{1.5929e-13} & \num{1.5931e-13} & \num{1.00} & \num{6.3719e-13} & \num{6.3999e-13}\\

\hline \hline

    \end{tabular}
    }
    \caption{Thermal reaction rate coefficients $k(T)$,
    in cm$^3$\,molecule$^{-1}$\,s$^{-1}$, calculated at different temperatures $T$,
    with and without a barrier correction due to the diagonal
    Born--Oppenheimer correction (DBOC)
    \cite{03MiPeScGa}
    for the reaction H$_2$ + D $\rightarrow$ HD + H.
    }
    \label{ArtTab:kT}
\end{table}

\subsection{Thermal reaction rate coefficients}

The extensive computations of this study have resulted in thermal
reaction rate coefficients $k(T)$ for the \ce{H2 + D -> HD + H} reaction
in the temperature range of $75-800$~K.
All $k(T)$ values shown in Table~\ref{ArtTab:kT}
have been computed with the CCI PES.\cite{02MiGaPe}
For half of the reported results, a DBOC-related correction \cite{03MiPeScGa}
was applied; these $k(T)$ values can be found under the heading
`with barrier correction'.
The sets of thermal reaction rate coefficients given in Table~\ref{ArtTab:kT}
are as follows:
(a) 
separate \textit{ortho} and \textit{para} contributions to the thermal reaction rate,
$k^{\textit{ortho}}(T)$ and $k^{\textit{para}}(T)$, respectively;
(b) the weighted sum,
$k^{\rm full}(T) = 3\,k^{\textit{ortho}}(T) + k^{\textit{para}}(T)$,
is the reaction rate coefficient that takes into account the contributions
from both nuclear spin states (assuming equilibrium between the nuclear spin
isomers at all temperatures considered);
and
(c) the set of 
extrapolated reaction rate coefficients, $k^{\rm extrapolated}$.
$k^{\textit{ortho}}(T)$, $k^{\textit{para}}(T)$, and $k^{\rm full}(T)$
contain only the directly computed $k^J(T)$ contributions up to $J_{\rm max}$. 
In clear contrast, $k^{\rm extrapolated}$
is obtained by including contributions up to $J=30$, in the sum of Eq. \eqref{eq:ktrot}.
The $k^J(T)$ values, where $J>J_{\rm max}$ are obtained by a linear extrapolation (Fig.~\ref{ArtFig:kJ}, \textit{vide infra}).
The linearity relations observed within the computed $k^J(T)$ values,
which is utilized during the extrapolation, will be discussed later.
Furthermore, Table~\ref{ArtTab:kT} gives the ratio of 
$k^{\textit{para}}(T)/k^{\textit{ortho}}(T)$, which helps illustrate the effect of nuclear spin isomerism on the reaction rate coefficients
as a function of temperature.
This ratio is different from 1.00 only below 150~K
and increases toward lower temperatures.
%
%
The comparison of the $k^{\rm full}$ and $k^{\rm extrapolated}$ values
in Table~\ref{ArtTab:kT} serves as a simple convergence test for the choices made for $J_{\rm max}$ (Table \ref{ArtTab:compparam}).
Finally, note that two sets of reaction rate coefficients are tabulated 
in Table~\ref{ArtTab:kT} for $T=100$ K:
one from a direct computation at $T=100$ K, and one derived from a computation
at $T=200$ K, using Eq.~\eqref{ArtEq:halftemp} during the evaluation of the
flux autocorrelation function, denoted as `100 = 200/2'.
The agreement of the two values proves the validity of Eq.~\eqref{ArtEq:halftemp}.

The computed $k(T)$ values obtained with or without the barrier 
correction \cite{03MiPeScGa} are significantly different at the lowest temperatures.
A similar result was obtained by Mielke and co-workers,\cite{03MiPeScGa}
who listed a difference of 34\% for their $k(T)$ values calculated at 167~K.
At lower temperatures, the correction is even larger;
at 75 K, the correction becomes a factor of 2.5.
The employed simple barrier correction formula assumes that
the reaction rate coefficient is proportional to $e^{-\beta \Delta E_{\rm b}}$,
where $\Delta E_{\rm b}$ is the correction to the barrier height.\cite{03MiPeScGa}
At the lowest temperatures, this assumption seems to fail, suggesting that 
more sophisticated 
schemes should be used to include the DBOC effect.
The effect of DBOC on the dynamics of the D + H$_2(v=4,j=0)$ reaction was 
studied in Ref.~\citenum{26FaHuZh}.
This study is not directly relevant to the present paper, 
nevertheless, it shows that for such light systems, the DBOC correction is not negligible.
Other works worth mentioning investigated effects due to the conical intersection between the two lowest electronic states and
the geometric phase for the H$_2$ + H reaction and its isotopic analogues.\cite{93WuKu,03Kendrick_2,03JuAl}



\begin{figure}[b!]
    \centering
    \includegraphics[width=1.0\linewidth]{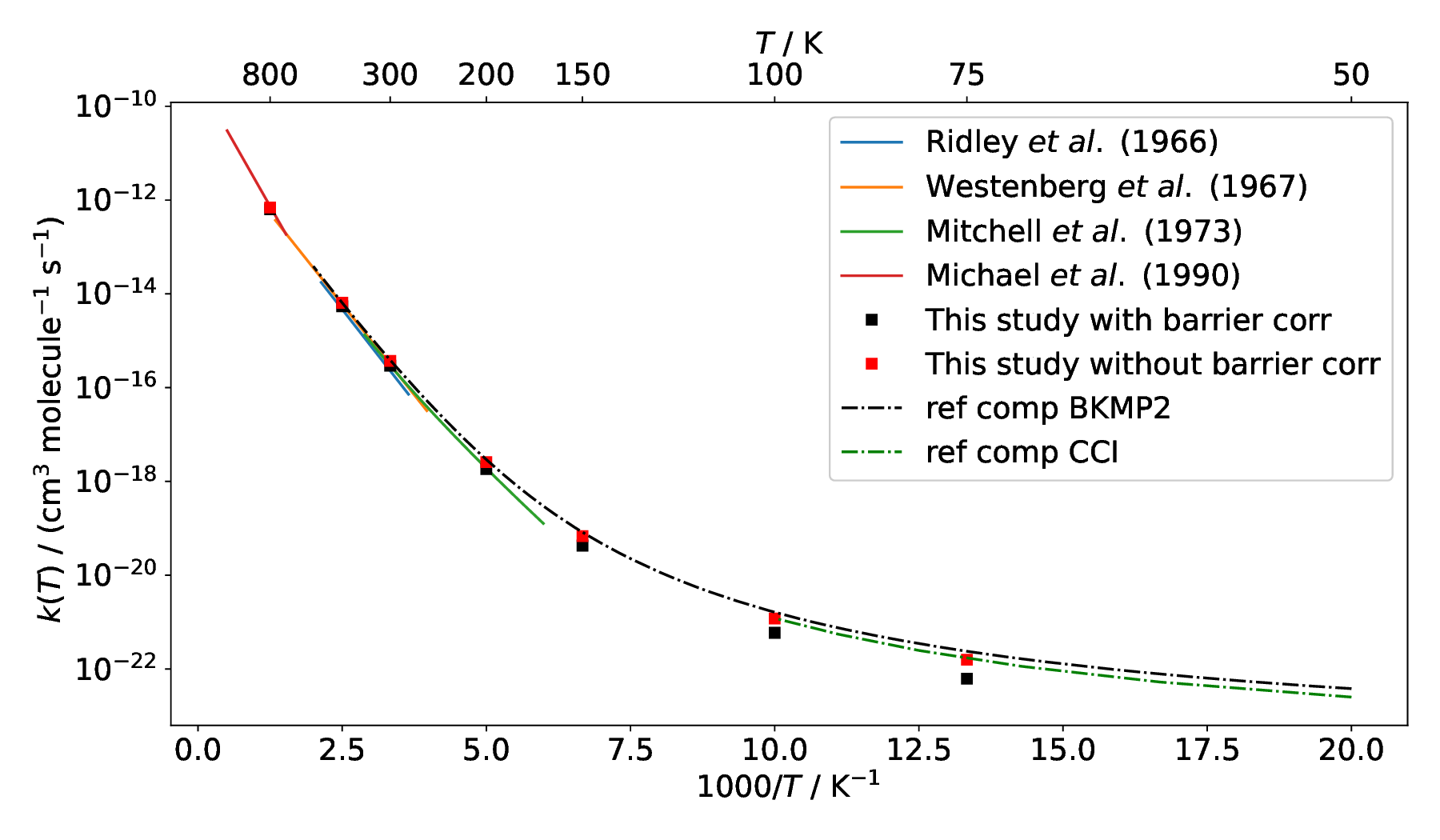}
    \caption{Arrhenius plot of the computed thermal reaction rate coefficients with and without the barrier correction, 
    along with measured \cite{66RiScLe,67WeHa,73MiRo,90MiFi} values (solid lines), 
    and literature data \cite{17RaGh} computed on the BKMP2 and  CCI PESs (dash-dotted lines).
    The measured data are plotted as the Arrhenius fits of Ref.~\citenum{03MiPeScGa}.
    }
    \label{ArtFig:arrhenius}
\end{figure}



Figure~\ref{ArtFig:arrhenius} contains Arrhenius plots of the computed
thermal reaction rate coefficients of this study, along with
measured \cite{66RiScLe,67WeHa,73MiRo,90MiFi} and computed \cite{17RaGh} values.
Clearly, and not surprisingly, the results of this study are in good agreement
with the experimental values from all four sources and with earlier theoretical results.
For a reaction studied from very low (near 0~K) to very high temperatures,
the Arrhenius plot, that is ln\,$k(T)$ \textit{vs}. $1/T$, is expected
to show the following shape: 
straight at high temperatures (no observable effect from tunneling),
flat at very low temperatures (when the reaction proceeds strictly due to tunneling,
the rate is independent of the temperature for exothermic reactions),
and curved in between.
Some of this expected behavior can be observed in Fig.~\ref{ArtFig:arrhenius},
even in the slopes of the experimental measurements conducted at different
temperature ranges.
It is also clear that the tunneling regime is not approached closely at 75~K,
the lowest temperature point of this study. 
As estimated in Ref.~\citenum{87TaMaNaOk}, for the \ce{H2 + D -> HD + H} reaction,
the flat regime should be visible below about 10~K.




While the computed $k(T)$ values are of primary interest for reaction kinetics,
it is equally important, from methodological and computational points of view,
to discuss some of the intermediate computational results leading to the thermal
reaction rate coefficients. 
These important aspects of the present study are elaborated in the next subsections.

\begin{figure}[t!]
    \centering
    \begin{subfigure}[b]{0.45\textwidth}
        \centering
        \includegraphics[trim={20cm 5cm 20cm 5cm},clip,width=1\linewidth]{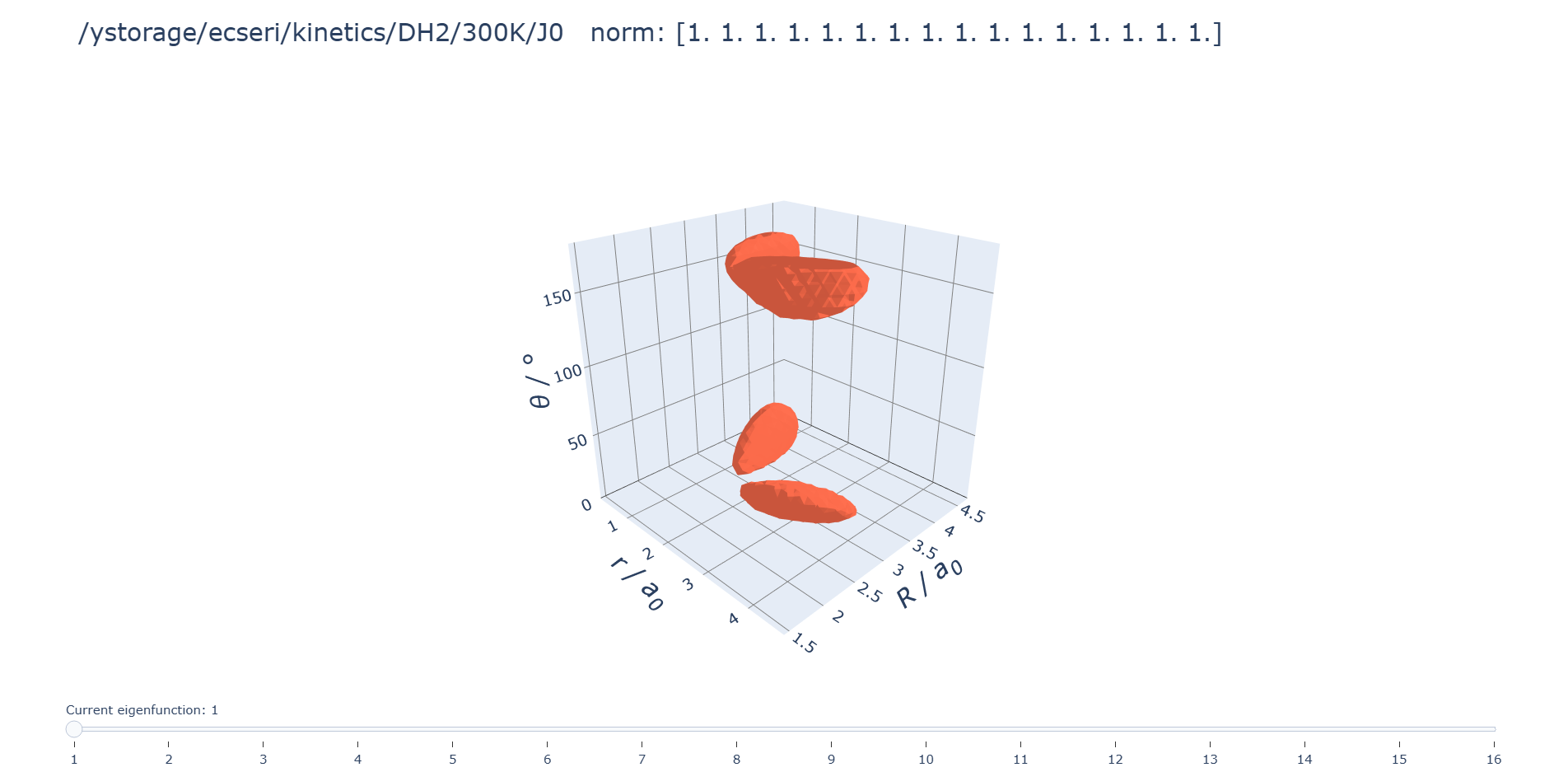}
        \caption{First eigenvector}
        \label{fig:sub1}
    \end{subfigure}
    \begin{subfigure}[b]{0.45\textwidth}
        \centering
        \includegraphics[trim={20cm 5cm 20cm 5cm},clip,width=1\linewidth]{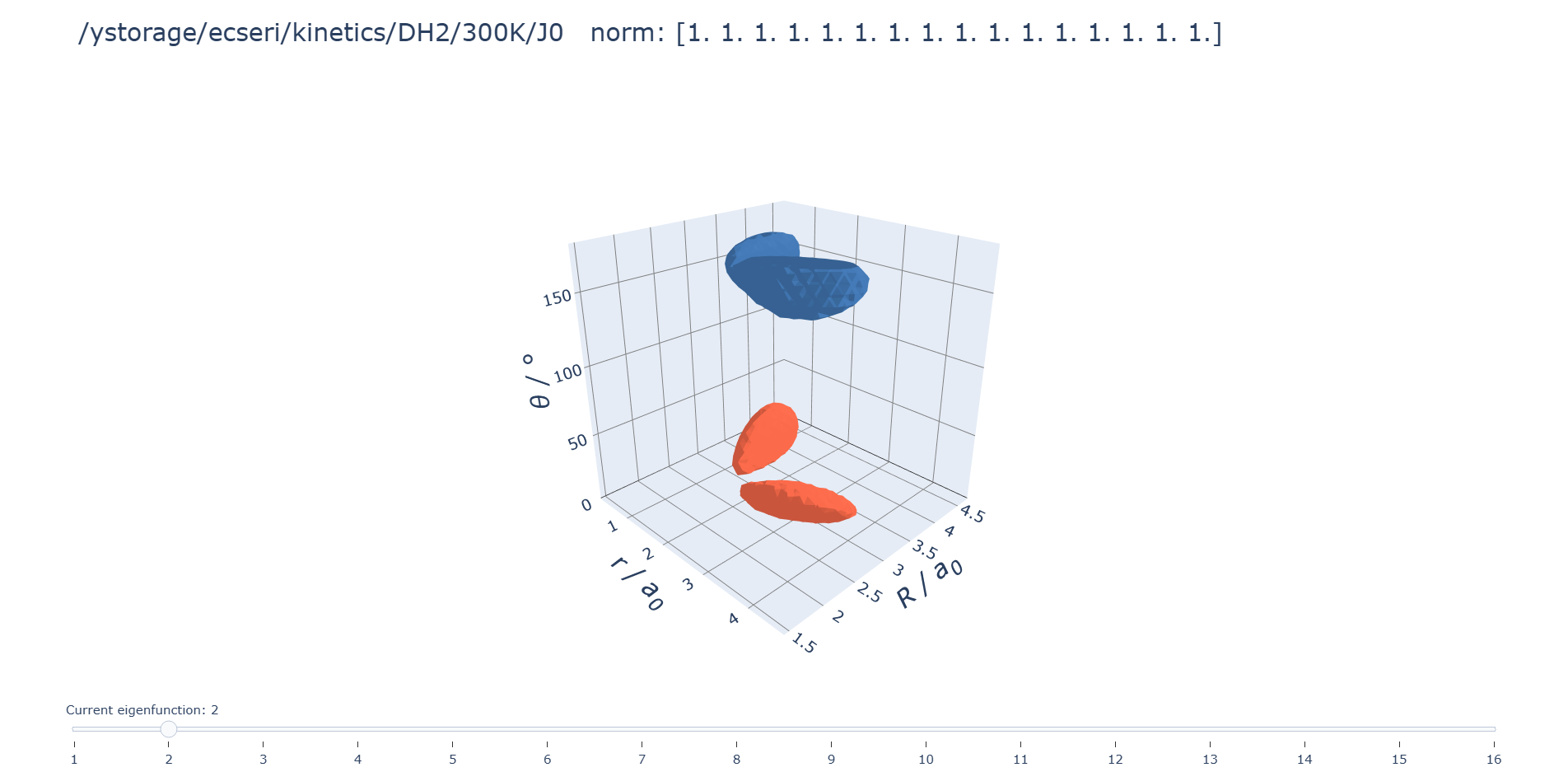}
        \caption{Third eigenvector}
        \label{fig:sub2}
    \end{subfigure}
    
    
    \begin{subfigure}[b]{0.45\textwidth}
        \centering
        \includegraphics[trim={20cm 5cm 20cm 5cm},clip,width=1\linewidth]{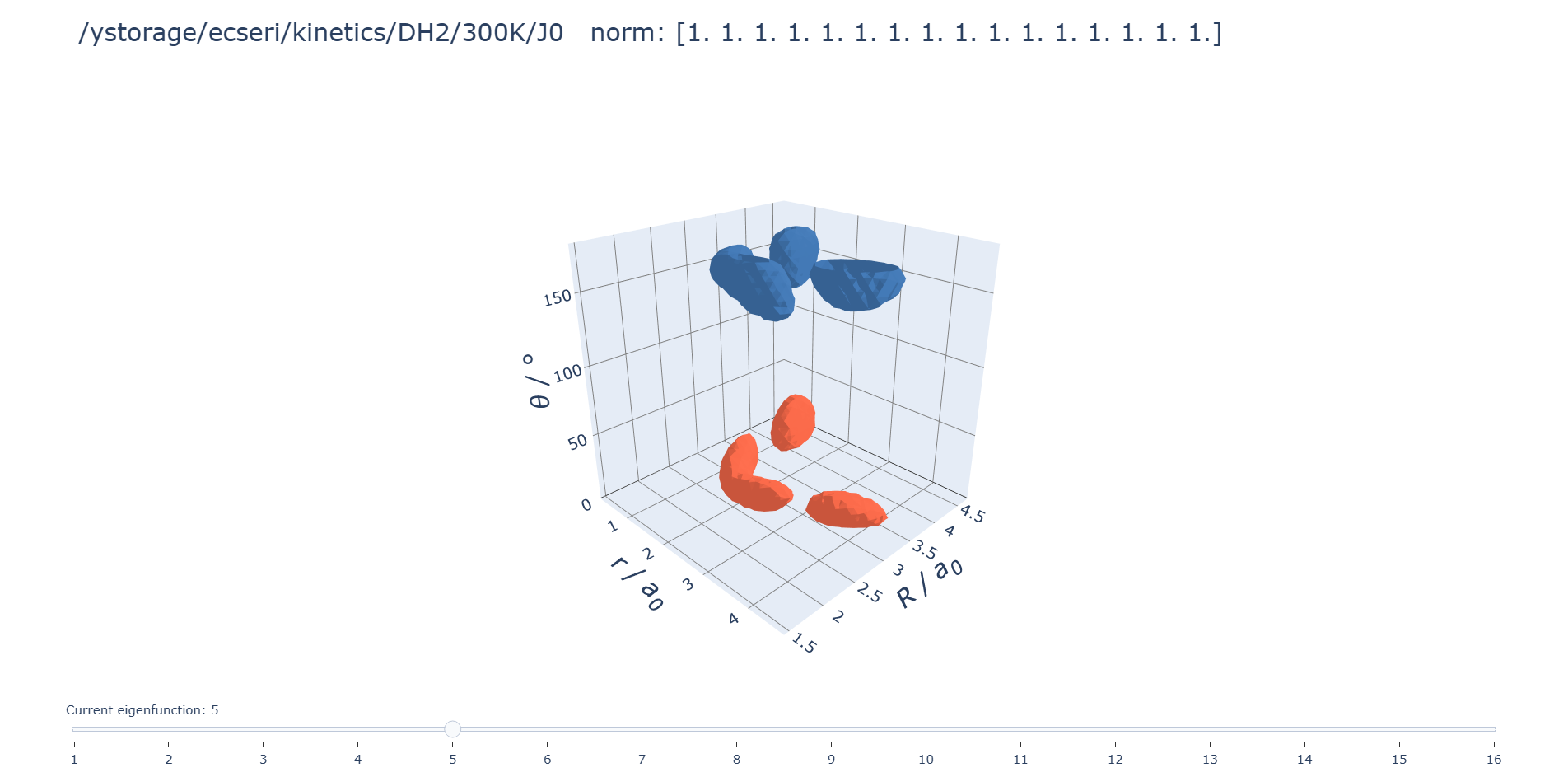}
        \caption{Fifth eigenvector}
        \label{fig:sub3}
    \end{subfigure}
    \begin{subfigure}[b]{0.45\textwidth}
        \centering
        \includegraphics[trim={20cm 5cm 20cm 5cm},clip,width=1\linewidth]{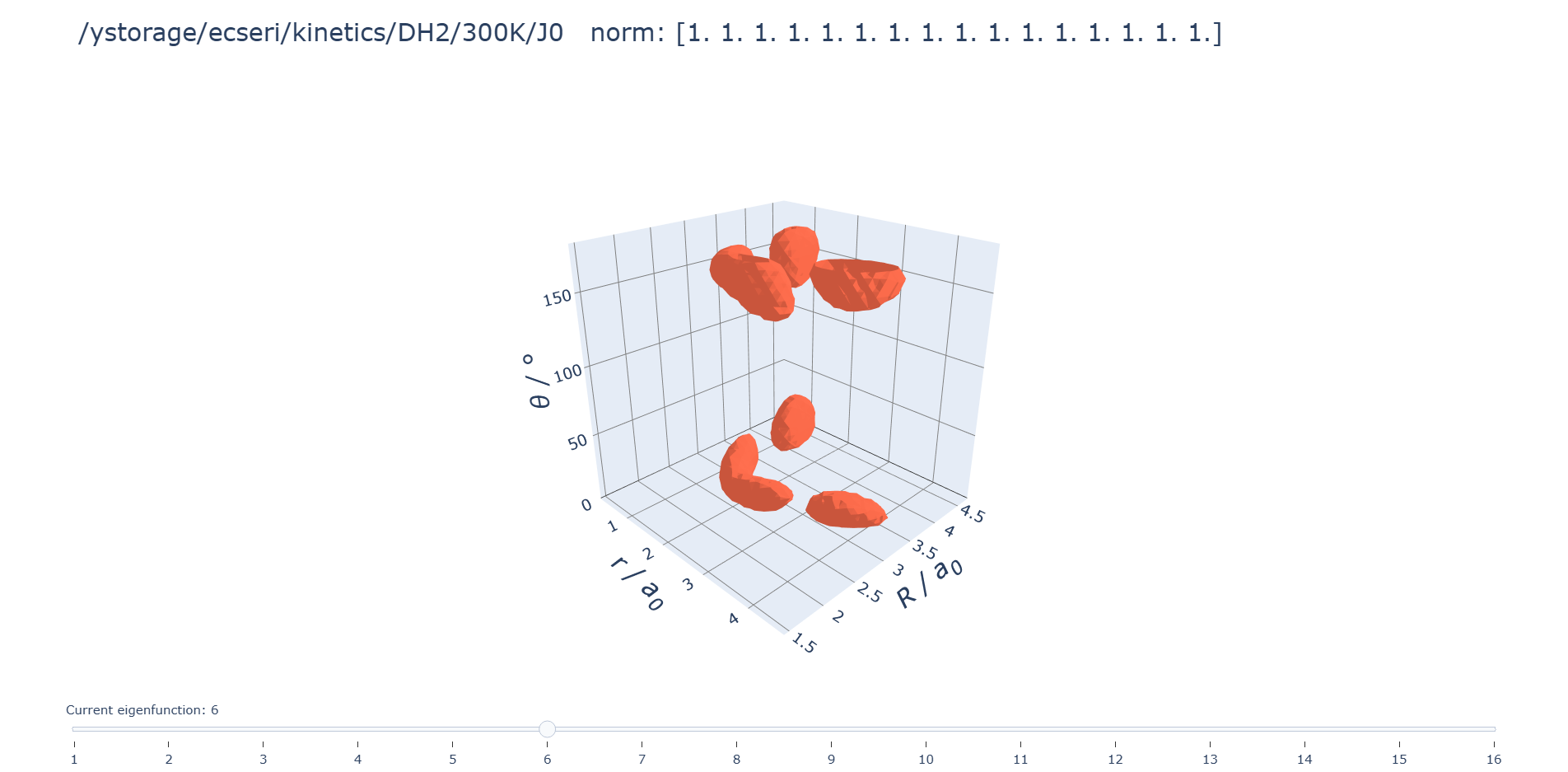} 
        \caption{Seventh eigenvector}
        \label{fig:sub4}
    \end{subfigure}
    \caption{The isosurface plot of the squared absolute value of the
    first (panel (a)), third (panel (b)), fifth (panel (c)), and seventh (panel (d)) thermal flux eigenvectors for the $J=0$
    \ce{H2 + D -> HD + H} reaction, at $T=300$~K. 
    Only one of the eigenvectors is plotted
    for each complex conjugate eigenvector pair.
    The coloring represents the symmetry of the eigenvectors under the exchange
    of the two hydrogen atoms.
    The first and seventh eigenvectors are symmetric, while the third and fifth
    eigenvectors are antisymmetric under the
    $\theta \rightarrow \pi - \theta$ transformation, that is, the permutation of the
    two hydrogen atoms.
    }
    \label{ArtFig:eigvecs}
\end{figure}

\subsection{Thermal flux eigensystem }

\begin{table}[b!]
    \centering
    \begin{tabular}{ccccc} \hline \hline
    Index & Symm.&150 K &  300 K & 800 K \\ \hline
 1,2  &even& $\pm1.4943 \times 10^{-19}$ &  $\pm2.4363\times 10^{-13}$  & $\pm9.8967\times 10^{-8}$  \\
 3,4  &odd& $\pm1.3296\times 10^{-19}$ &  $\pm2.4296\times 10^{-13}$  & $\pm9.8967\times 10^{-8}$  \\
 5,6  &odd& $\pm7.4917\times 10^{-21}$ &  $\pm8.6659\times 10^{-15}$  & $\pm9.5108\times 10^{-9}$  \\
 7,8  &even& $\pm8.7008\times 10^{-21}$ &  $\pm8.6796\times 10^{-15}$  & $\pm9.5114\times 10^{-9}$  \\ \hline \hline
    \end{tabular}
    \caption{The first eight thermal flux ($J=0$) eigenvalues, 
    given in atomic units (a.u.),
    at three different temperatures. 
    The column Symm. specifies the symmetry of thermal flux eigenstates under the permutation of the two hydrogen atoms.
    }
    \label{ArtTab:eigvals}
\end{table}


Due to the fact that the thermal flux operator is purely imaginary and Hermitian,
$\hat{F}(\beta)^* = -\hat{F}(\beta)$, 
its eigenvalues come in $\pm$ pairs, and the corresponding eigenvectors
are complex conjugates of one another.
The first few highest (in absolute value) thermal flux eigenvalues of the
\ce{H2 + D -> HD + H}
reaction with the $J=0$ constraint are given in Table~\ref{ArtTab:eigvals}
at a few selected temperatures.
The most notable qualitative feature of the data is that, at higher temperatures,
quasidegenerate pairs of the $\pm$ pairs start to appear,
yielding eigenvalue quadruplets with close-lying absolute values.
This can be explained by the fact that at higher temperatures the two reaction channels
become virtually independent.
The quasidegeneracy is lost at lower temperatures.



\begin{figure}[t!]
    \centering
    \begin{subfigure}[b]{0.49\textwidth}
        \centering
        \includegraphics[trim={20cm 5cm 20cm 5cm},clip,width=1\linewidth]{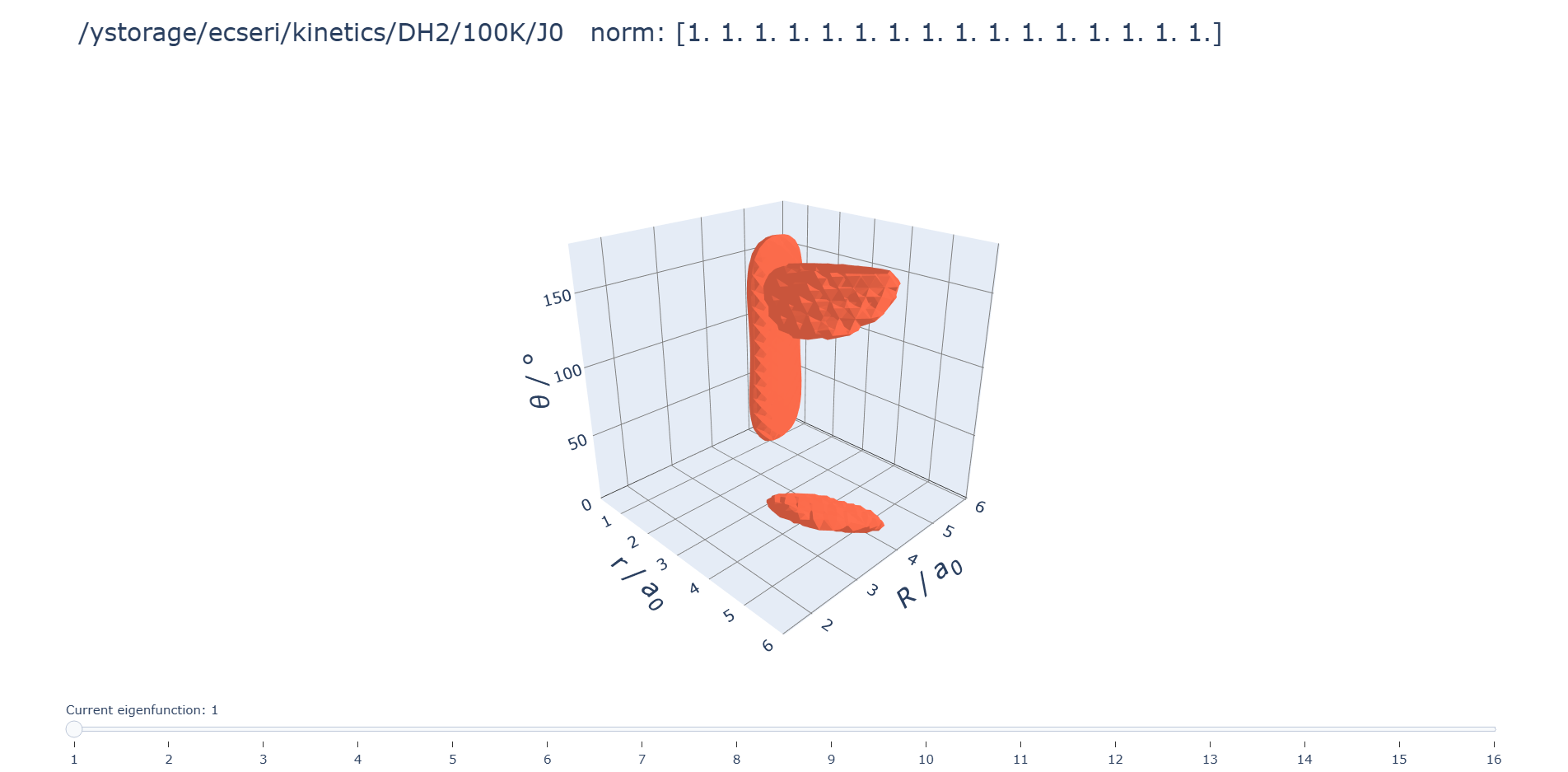}
        \caption{100 K}
    \end{subfigure}
    \begin{subfigure}[b]{0.49\textwidth}
        \centering
        \includegraphics[trim={20cm 5cm 20cm 5cm},clip,width=1\linewidth]{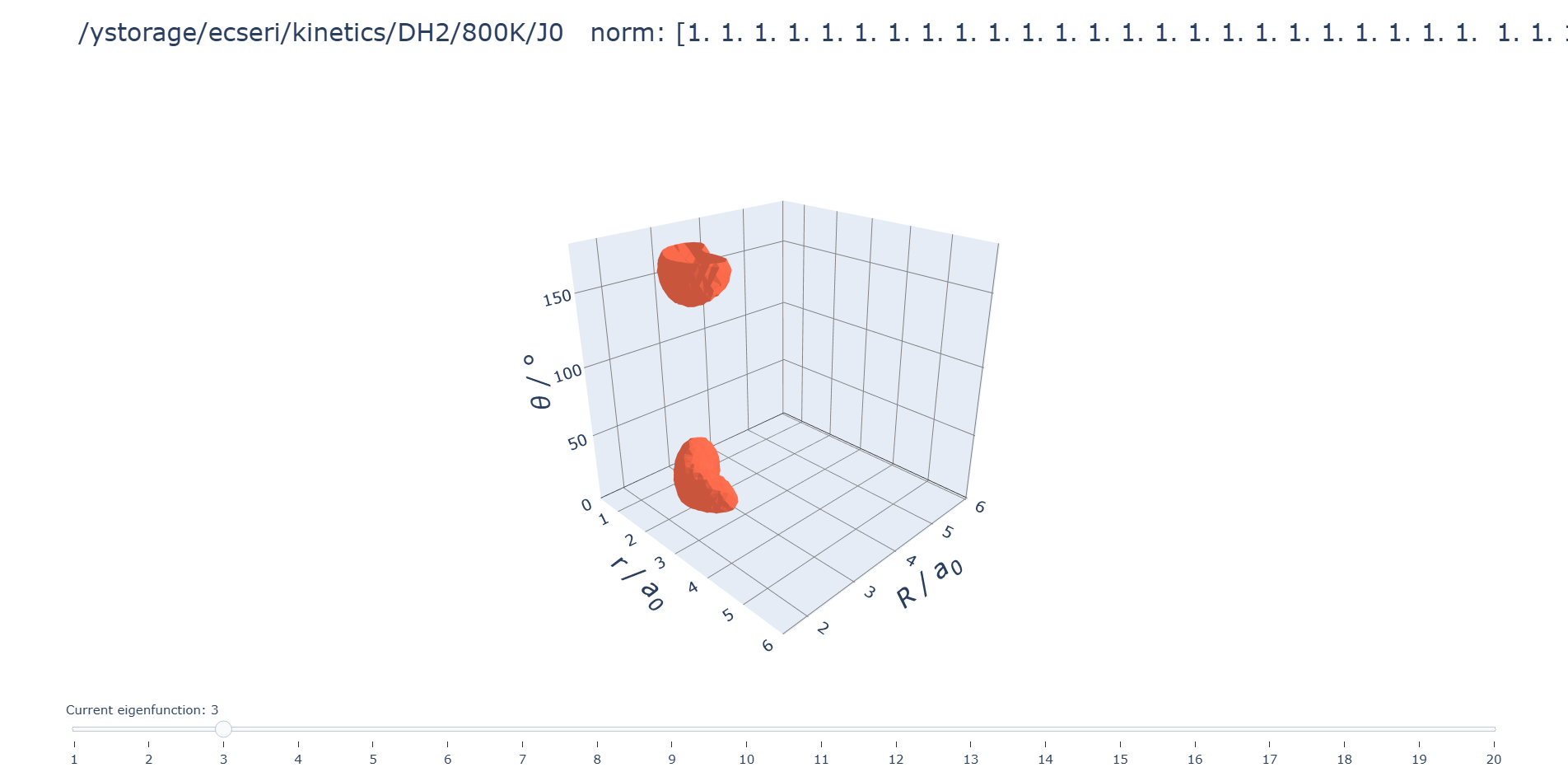}
        \caption{800 K}
    \end{subfigure}
    \caption{The isosurface plot of the squared absolute value of the
    first eigenstate (symmetric under the exchange of the two hydrogen atoms, \textit{para}) 
    of the thermal flux at two different temperatures
    (panel (a): $T=100 ~ \textrm{K}$, panel (b): $T=800 ~ \textrm{K}$).
    }
    \label{ArtFig:TFE100K}
\end{figure}

The squared absolute value of the thermal flux $J=0$ eigenvectors at $T=300$~K
is plotted in Fig.~\ref{ArtFig:eigvecs}. 
The two colors in these plots represent the symmetry of the
eigenvectors under the permutation
of the two indistinguishable hydrogen atoms. 
The first and seventh eigenvectors, Figs.~\ref{ArtFig:eigvecs}(a)
and \ref{ArtFig:eigvecs}(d), respectively, are symmetric;
thus, these are associated with \textit{para} states.
The third and fifth eigenvectors, Figs.~\ref{ArtFig:eigvecs}(b)
and \ref{ArtFig:eigvecs}(c), respectively, are antisymmetric;
thus, these two are \textit{ortho} states. 
At higher temperatures, the \textit{ortho} and \textit{para} reaction rate coefficients
are virtually equal due to quasidegeneracies in the spectrum of the thermal flux. 
At lower temperatures, the \textit{ortho} and \textit{para} reaction rate
coefficients differ and the quasidegeneracy is lost.
It is important to note, however, that the quasidegeneracy of the thermal flux eigenvalues alone is not a quantitative measure of the $k^\textit{para}/k^\textit{ortho}$ ratio. 
For example, while the quasidegeneracy is already lifted at $T=150$ K (see Table \ref{ArtTab:eigvals}), the $k^\textit{para}/k^\textit{ortho}$ ratio is still 1.00. 
On the other hand, at $T=75$ the ratio is 1.53, although the rate coefficients were computed using the $T=150$ K thermal flux eigenpairs and Eq. \eqref{ArtEq:halftemp}.

Further insight into the thermal flux eigenstates can be gained from
Fig.~\ref{ArtFig:TFE100K}, which shows the first thermal flux eigenstate
at two different temperatures, $T=100$~K and $T=800$~K.
Clearly, the picture is rather different at the two temperatures.
As expected, thermal flux eigenstates are more localized in the vicinity of the
transition state at $T=800$ K,
whereas at $T=100$ K, they are located farther away from the transition state,
in the reactant and product valleys.
Note that the transition state is located at $r=1.7572\ a_0$, $R=2.6358\ a_0$,
and $\theta = 0^\circ$ and $180^\circ$ for the two reaction channels.
Since the barrier between the two channels is high at the transition state,
at high temperatures the thermal flux eigenstates 
can be divided into two non-interacting parts, each for the two reaction channels,
causing the quasidegeneracy in the thermal flux eigenvalues (see Table~\ref{ArtTab:eigvals}).
However, at distances far from the transition state in the reactant channel,
the barrier between the two channels decreases.
Therefore,
as the thermal flux eigenstates leave the transition state region as
the temperature decreases, 
they reach regions where the barrier is no longer high enough to prevent
the interaction between the two channels; thus, 
the degeneracy is lifted in the eigenvalues and the 
\textit{ortho} and \textit{para} reaction rate coefficients eventually start to differ.



\subsection{Flux autocorrelation functions}

\begin{figure}[t!]
    \centering
    \includegraphics[width=0.95\linewidth]{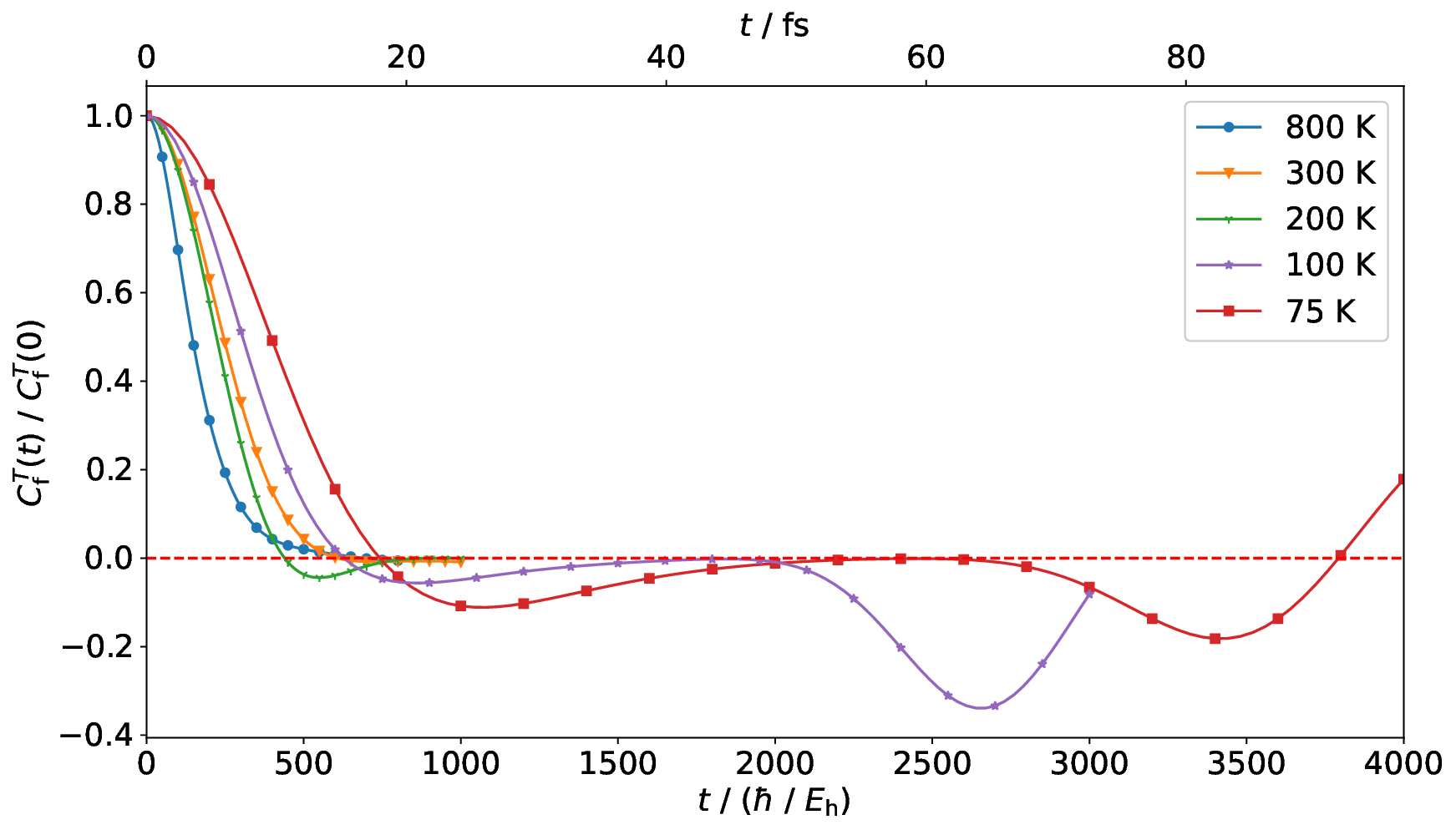}
    \caption{The flux autocorrelation function summed for all relevant $J$ and nuclear spin states, normalized by its value at $t=0$, and
    plotted for a few selected temperatures as a function of time.
    All flux autocorrelation functions converge to zero; thus, an appropriate integration upper bound can be chosen. 
    The diverging tails, see, \textit{e.g.}, 
    $t>2000\ \hbar/E_h$ for $T=100$~K, 
    are caused by unphysical reflection from the end of the DVR grid.
    }
    \label{ArtFig:cff}
\end{figure}

Figure~\ref{ArtFig:cff} shows the flux autocorrelation function
$C_{\rm f}^T(t)$ at a few selected temperatures.
The functions are normalized by their value at $t=0$;
otherwise, they would differ by several orders of magnitude.
All functions converge to zero eventually; however, the autocorrelation functions
at lower temperatures converge at a slower rate, as expected. 
At certain temperatures, diverging tails can be observed. 
These unphysical parts of the flux autocorrelation function
are caused by the reflection of the thermal flux eigenstates
from the end of the grid during the real time evolution. 
A complex absorbing potential (CAP) could avoid such reflections; 
however, in the current study, we avoided the use of a CAP
since all the flux autocorrelation functions converged to zero
before the reflections emerged.


\subsection{Rotational effects}

    


\begin{figure}[t!]
    \centering
    \includegraphics[width=0.95\linewidth]{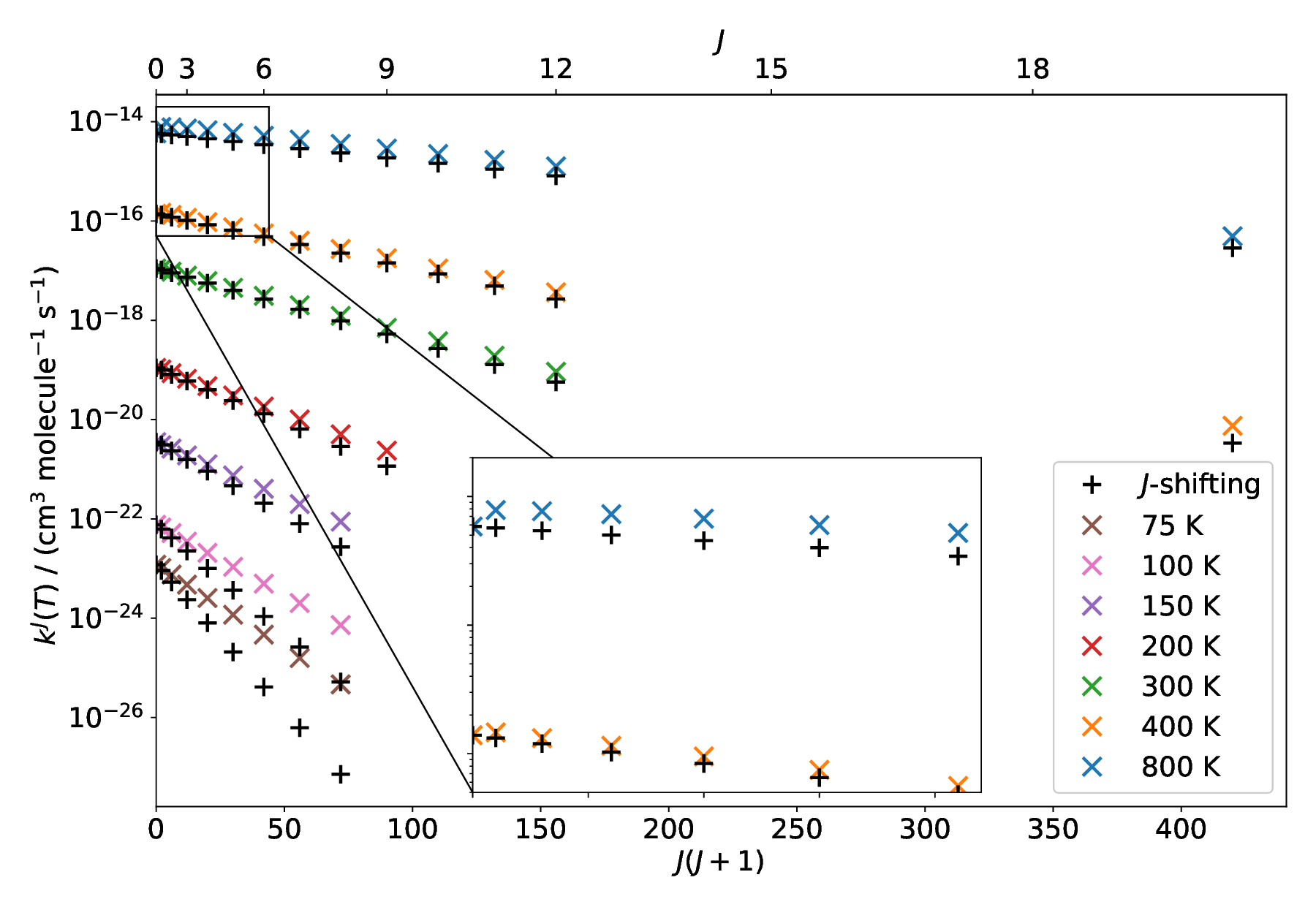}
    \caption{The semilog plot of the $k^J(T)$ contributions to the full $k(T)$ value as a function of
    $J(J+1)$ at temperatures between 75 and 800~K. 
    Thermal rate coefficients obtained by the $J$-shifting approximation are also shown. 
    The deviations observed at low temperatures are caused by the change of
    the effective rotational constant,
    while at high temperatures they can be attributed
    to non-negligible $K\neq0$ contributions for low $J$ values (see the inset).
    }
    \label{ArtFig:kJ}
\end{figure}

The different computed $k^J(T)$ contributions are plotted in Fig.~\ref{ArtFig:kJ}
as a function of $J(J+1)$ at all the temperatures chosen for the
$k(T)$ computations. 
The rotational contributions computed using the $J$-shifting approximation
are also shown. 
Within the $J$-shifting approximation, $k^J(T)$ can be obtained as
\begin{equation}
    k^J(T) =  k^{J=0}(T)\,e^{-\beta E_J^\ddagger} ,
\end{equation}
where $E_J^\ddagger$ is the rigid-rotor rotational energy of the reactive system
at the transition-state geometry. 
Since, in the case of the \ce{H2 + D -> HD + H} reaction,
the transition-state structure is linear, 
the energy level formula $E_J^\ddagger = B^\ddagger J(J+1)$ applies with $B^\ddagger = 7.04 ~ \textrm{cm}^{-1}$.
Therefore, the $\log k^J(T) - J(J+1)$ plot is linear within the
$J$-shifting approximation (see Fig.~\ref{ArtFig:kJ}).
Remarkably, exactly computed $k^J(T)$ values show a similar behavior,
deviating from linearity only at high temperatures for low $J$ values (see the inset in Fig.~\ref{ArtFig:kJ}), 
due to non-negligible $K\neq0$ contributions.
As is clear from Fig.~\ref{ArtFig:kJ},
the slopes of exact $k^J(T)$ values and the $J$-shifting ones are virtually equal
at high temperatures, while they deviate at low temperatures. 
The exact values of the slopes are tabulated in Table~\ref{ArtTab:slopes}. 
Effective rotational constants $B^{\rm eff}$ can be computed from the slope of the exact $k^J(T)$ values for each temperature,
these rotational constants can also be found in Table~\ref{ArtTab:slopes}.
At higher temperatures, $B^{\rm eff}$ is in good agreement with $B^\ddagger$.
As the temperature is reduced, $B^{\rm eff}$ decreases appreciably and as a consequence, the deviation 
between $B^{\rm eff}$ and $B^\ddagger$ increases.
These findings suggest that the effective linear transition-state structure consistent with $B^{\rm eff}$ is temperature dependent. 
Moreover, a decrease in $B^{\rm eff}$ implies that the bonds of the effective transition-state structure are elongated at lower temperatures.
We believe that this observation is in line with the fact that  thermal flux eigenstates leave the transition state region as the temperature is reduced (see Fig.~\ref{ArtFig:TFE100K}).

\begin{table}[t!]
    \centering
    \begin{tabular}{c|cll} \hline \hline
         $T$ / K & $B^{\rm eff} / $\ \cm{} & fitted & $J$-shifting \\ \hline
         75  & 3.98 & $-0.07632$ & $-0.13499$ \\
         100 & 4.43 & $-0.06380$  & $-0.10124$ \\
         150 & 5.31 & $-0.05097$ & $-0.06749$ \\
         200 & 5.87 & $-0.04226$ & $-0.05062$ \\
         300 & 6.43 & $-0.03082$ & $-0.03375$ \\
         400 & 6.52 & $-0.02346$ & $-0.02531$ \\
         800 & 6.83 & $-0.01229$ & $-0.01266$ \\ \hline  \hline
    \end{tabular}
    \caption{Slopes of the curves in Fig. \ref{ArtFig:kJ} obtained by
    fitting a line to the last two data points
    as well as computed using the $J$-shifting approximation, and effective rotational constants derived from the fitted slopes.
    }
    \label{ArtTab:slopes}
\end{table}

\section{Conclusions}
Ever since the ground-breaking, classic studies of 
Eyring, Evans, and Polanyi,\cite{31EyPo,35Eyring,35EvPo}
the various statistical, classical, 
semiclassical, and quantum approaches developed have helped
explain several characteristic observations in chemical kinetics, among them
(a)~general trends in the preexponential factors of the Arrhenius equation,
(b)~kinetic isotope effects of bimolecular reactions,
(c)~the temperature dependence of thermal reaction rate coefficients,
(d)~the role of quantum effects (zero-point vibrations, tunneling, 
identical particle symmetry, resonances, interference, the still somewhat exotic
Aharonov--Bohm effect \cite{59AhBo}),
and (e)~deviations of certain reactions even from the generalized Arrhenius form
of the rate equation.
One of the principal aims of the present study was to provide an improved implementation of
a direct quantum mechanical algorithm based on flux autocorrelation, applicable
for the efficient computation of thermal reaction rate coefficients.
The methodology was tested on the prototypical \ce{H2 + D -> HD + H} reaction
over the wide temperature range of $75-800$~K.

There are several characteristics of the improved methodology worth emphasizing.
First, it is the strong belief of the authors
    that the combination of the fully numerical formulation of the
    rovibrational Hamiltonian allowed within the \texttt{GENIUSH} approach with
    the `direct' method, originally developed by Yamamoto, Miller, and Manthe,
    provides a unique opportunity to develop a
    black-box-type approach for the computation of thermal reaction rate coefficients.
\texttt{GENIUSH} can construct the exact and general rovibrational Hamiltonian
for arbitrary molecules, internal coordinates, and body-fixed frame embeddings with little effort
on the user's side; most importantly, no analytical form of the 
kinetic energy operator is needed.
Thus, the \texttt{GENIUSH} approach is useful not only for obtaining
time-independent rovibrational energy levels and eigenstates,\cite{12CsFaSzMa}
but also for describing the rate of chemical reactions.
It should also be mentioned that \texttt{GENIUSH} has already been used for a
time-dependent quantum dynamics study.\cite{19FaMaCs}
Second, the flexibility of choosing any internal coordinates and embeddings in \texttt{GENIUSH} facilitates symmetry considerations for equivalent nuclei,
which can reduce the computational cost and help enforce the Pauli exclusion principle.
Third, reduced-dimensional models are easy to implement in \texttt{GENIUSH}, which can find application for the quantum mechanical description of larger systems,
especially if there are internal degrees of freedom that are less relevant for the chemical reaction under investigation.
     
In the current work, \texttt{GENIUSH} has been applied to the \ce{H2 + D -> HD + H} reaction,
performing numerically exact full-dimensional computations that include rotational and nuclear spin degrees of freedom as well.
Due to numerical issues experienced with the Podolsky form of the general rovibrational kinetic energy operator used extensively in earlier work,
we had to devise a new form of the kinetic energy (termed the partially rearranged form) to obtain correct reaction rate coefficients.
The \ce{H2 + D -> HD + H} reaction, studied both experimentally and theoretically in great detail, provided an outstanding testing ground for our algorithm.
Excellent agreement with earlier experimental and computational results has been achieved by the current work in a broad temperature range for the \ce{H2 + D -> HD + H} reaction.
In addition, we have also presented interesting results related to rotational effects and nuclear spin (such as the temperature dependence of the ratio $k^\textit{para}/k^\textit{ortho}$).

We envision that the algorithm and the code developed and described in this paper will allow for the computation of thermal reaction rate coefficients for significantly larger systems than the triatomic test system examined here.
One possible direction is the investigation of chemical reactions that cannot accurately be described
by the most commonly used classical and semiclassical methods due to the
importance of quantum effects.
Such effects are particularly relevant in reactions involving light atoms
and occurring at low temperatures -- 
conditions commonly encountered in astrochemical environments.

\appendix

\section{Appendix: Partially rearranged form of the exact
kinetic energy operator}

In most computations of rovibrational energy levels and eigenstates utilizing the
\texttt{GENIUSH} code,\cite{09MaCzCs,11FaMaCs} the Podolsky form \cite{28Podolsky}
of the exact kinetic energy operator has been used.
The most time-consuming part of a rovibrational computation is the
determination of the required energy levels and eigenstates of the rovibrational Hamiltonian
matrix $\mathbf{H}$,
which is carried out by the iterative Lanczos algorithm \cite{50Lanczos}
in \texttt{GENIUSH}.

To this end, the matrix representation of the Hamiltonian
is set up in a multidimensional direct-product DVR basis,
and matrix-vector products $\mathbf{H} \mathbf{v}$ are evaluated
by an efficient algorithm that exploits the sparsity of $\mathbf{H}$.
$\mathbf{H}$ is constructed by inserting resolutions of identity between the different operators in the Podolsky form
of the kinetic energy [see Eq. \eqref{eq:tvibpod}].
This procedure provides numerically exact bound rovibrational energy levels
and wave functions.
Unfortunately, serious numerical issues arise if the Podolsky form
is applied to compute bimolecular reaction rate coefficients
using the resolution-of-identity approach.

For the sake of simplicity, let us assume that the quantities
$\tilde{g}$ and $G_{kl}$ do not depend on the internal coordinates,
and a single vibrational degree of freedom is considered.
In this case, the vibrational kinetic energy is equal to
\begin{equation}
    \hat{T}^\textrm{vib} = - \frac{\hbar^2}{2} \frac{\textrm{d}}{\textrm{d}q} G \frac{\textrm{d}}{\textrm{d}q},
\end{equation}
where $G$ is a constant and $q$ is the vibrational coordinate.
Using the resolution-of-identity approach, we get
\begin{equation}
    \mathbf{T}^\textrm{vib,res} = - \frac{\hbar^2}{2} G \mathbf{D}^\mathrm{(1)}\mathbf{D}^\mathrm{(1)},
\label{eq:D1}
\end{equation}
where $\mathbf{D}^\mathrm{(1)}$ is the matrix representation of the
first derivative operator.
Without inserting a resolution of identity between the two derivative operators,
one would get
\begin{equation}
    \mathbf{T}^\textrm{vib} = - \frac{\hbar^2}{2} G \mathbf{D}^\mathrm{(2)},
\label{eq:D2}
\end{equation}
where $\mathbf{D}^\mathrm{(2)}$ denotes the second derivative matrix.
The two approaches yield nearly identical results if $\mathbf{D}^\mathrm{(2)}$ is
well approximated by $\mathbf{D}^\mathrm{(1)}\mathbf{D}^\mathrm{(1)}$.

\begin{figure}[t!]
    \centering
    \includegraphics[width=0.65\linewidth]{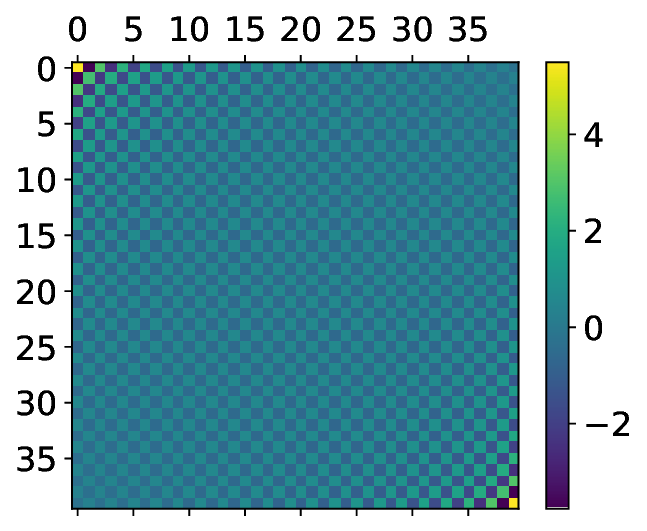}
    \caption{Graphical representation of the elements of the 
    $\mathbf{D}^\mathrm{(1)}\mathbf{D}^\mathrm{(1)} - \mathbf{D}^\mathrm{(2)} $
    matrix [see Eqs.~(\ref{eq:D1}) and (\ref{eq:D2})], based on 40 sine-DVR points.
    The differences become large (in absolute value) at the top left and bottom right corners
    of the difference matrix.
    Consequently, the resolution-of-identity approximation causes the largest errors at the boundaries of the grid.}
    \label{fig:DVRerror}
\end{figure}

The difference between $\mathbf{D}^\mathrm{(1)}\mathbf{D}^\mathrm{(1)}$
and $\mathbf{D}^\mathrm{(2)}$ is shown in Fig. \ref{fig:DVRerror} for
a basis of 40 sine-DVR functions.
The difference matrix has large elements (in absolute value) at its top left and bottom right corners,
corresponding to DVR points at the two edges of the grid.
Thus, the approximation 
$\mathbf{D}^\mathrm{(2)} \approx \mathbf{D}^\mathrm{(1)}\mathbf{D}^\mathrm{(1)}$
is inaccurate at the edges of the grid and accurate in the middle region.
If more DVR grid points are used, the central matrix elements of 
$|\mathbf{D}^\mathrm{(1)}\mathbf{D}^\mathrm{(1)} - \mathbf{D}^\mathrm{(2)}|$
become smaller, eventually approaching zero,
while the matrix elements at the edges never vanish, regardless of the grid size.
Such errors are irrelevant if bound states with zero amplitude at the edges
of the grid are considered.
However, if $\hat{F}(\beta)$ is to be diagonalized for a bimolecular reaction,
the edges of the grid are highly relevant.
Thus, the insertion of a resolution of identity between the derivative operators
causes severe numerical errors.

To obtain correct thermal reaction rate coefficients,
it is necessary to rearrange the rovibrational kinetic energy operator.
In the partially rearranged form of Eq. \eqref{ArtEq:prearr_deriv},
diagonal ($k=l$) terms in the vibrational kinetic energy operator,
where a resolution of identity would normally be inserted
between the two first-derivative operators $\frac{\partial}{\partial q_k}$, 
are rearranged so that, in the final form, a second-derivative operator appears
instead of two first-derivative operators.
A somewhat similar rearrangement was performed in Ref. \citenum{19AvMa};
however, it was applied to an already rearranged form
of the rovibrational kinetic energy operator and not to the Podolsky form.

The partially rearranged form of $\hat{T}^\textrm{vib}$ is Hermitian. However, depending on the numerical approach, 
the matrix representation of $\hat{T}^\textrm{vib}$
is not necessarily Hermitian.
In \texttt{GENIUSH}, resolutions of identity are inserted
into Eq. \eqref{ArtEq:prearr_deriv} between $\frac{\partial (G_{kk} \tilde{g}^{1/2})}{\partial q_k}$ and 
$\frac{\partial}{\partial q_k}$, and between $G_{kk} \tilde{g}^{1/2}$ and $ \frac{\partial^2}{\partial q_k^2}$.
Let us define $\mathbf{A}_{kk}$ and $\mathbf{A}_{kk}'$ as the matrix representations of
$G_{kk} \tilde{g}^{1/2}$ and $\frac{\partial (G_{kk} \tilde{g}^{1/2})}{\partial q_k}$ in DVR, respectively, 
and denote derivative matrices of the $k$th vibrational degree of freedom by $\mathbf{D}_k^{(1)}$ and $\mathbf{D}_k^{(2)}$. 
Using this notation, we get
\begin{gather}
    \left(\mathbf{A}_{kk}'  \mathbf{D}_k^{(1)} + \mathbf{A}_{kk} \mathbf{D}_k^{(2)}\right)^\dagger =   (\mathbf{D}_k^{(1)})^\textrm{T} (\mathbf{A}_{kk}')^\textrm{T} + (\mathbf{D}_k^{(2)})^\textrm{T} \mathbf{A}_{kk}^\textrm{T} \nonumber \\
    = (\mathbf{D}_k^{(1)})^\textrm{T} \mathbf{A}_{kk}' + (\mathbf{D}_k^{(2)})^\textrm{T} \mathbf{A}_{kk} 
    \neq \mathbf{A}_{kk}'  \mathbf{D}_k^{(1)} + \mathbf{A}_{kk} \mathbf{D}_k^{(2)},
\label{eq:mxnh}
\end{gather}
where the adjoint was replaced by the transpose since all matrices are real-valued, and 
$\mathbf{A}_{kk}$ and $\mathbf{A}_{kk}'$ are diagonal in a DVR basis, thus symmetric.
$\mathbf{D}_k^{(1)}$ and $\mathbf{D}_k^{(2)}$ are usually
antisymmetric and symmetric, respectively, 
for most DVR basis types; however, in general, this is not true.
The matrix representation of the kinetic energy operator should be Hermitian;
thus, Eq. \eqref{eq:mxnh} must be symmetrized as follows:
\begin{align}
    ( \mathbf{A}_{kk}' \mathbf{D}_k^{(1)} + \mathbf{A}_{kk} \mathbf{D}_k^{(2)} + (\mathbf{D}_k^{(1)})^\textrm{T} \mathbf{A}_{kk}' + (\mathbf{D}_k^{(2)})^\textrm{T} \mathbf{A}_{kk} )/2.
\end{align}
Note that such symmetrization is not needed in the off-diagonal terms ($k \ne l$)
of $\hat{T}_\textrm{vib}$,
since the two first derivative operators $\frac{\partial}{\partial q_k}$
and $\frac{\partial}{\partial q_l}$ act on different degrees of freedom.
In this case, the resolution of identity can be safely inserted between
$\frac{\partial}{\partial q_k}$ and $\frac{\partial}{\partial q_l}$.
The cost of matrix-vector multiplication is increased
because of the symmetrization.
However, since it is only needed in the diagonal ($k=l$) terms of 
$\hat{T}^{\textrm{vib}}$,
and the time complexity of the matrix-vector multiplication algorithm remains unchanged,
the cost increment caused by the symmetrization is manageable.
Thus, the final symmetrized matrix representation of the
partially rearranged operator $\hat{T}^\textrm{vib}$ reads as
\begin{gather}
    \mathbf{T}^\textrm{vib} = \frac{\hbar^2}{2} \sum_{k \neq l}^{D} \mathbf{\tilde{g}}^{-1/4} (\mathbf{D}_k^{(1)})^\textrm{T} \mathbf{A}_{kl} \mathbf{D}_l^{(1)} \mathbf{\tilde{g}}^{-1/4}
    - \frac{\hbar^2}{4} \sum_{k = 1}^{D} \mathbf{\tilde{g}}^{-1/4}  
    \Bigl( \mathbf{A}_{kk}' \mathbf{D}_k^{(1)} \nonumber \\ 
    + \mathbf{A}_{kk} \mathbf{D}_k^{(2)} 
     +(\mathbf{D}_k^{(1)})^\textrm{T} \mathbf{A}_{kk}' + (\mathbf{D}_k^{(2)})^\textrm{T} \mathbf{A}_{kk}
    \Bigr)
    \mathbf{\tilde{g}}^{-1/4}
\end{gather}
which is now implemented in \texttt{GENIUSH}.\\\\

\section{Acknowledgements}
This paper is dedicated to the late Professor John F. Stanton,
a friend and colleague with whom the second author (AGC) 
has enjoyed numerous stimulating discussions about theoretical
and computational chemistry, resulting in several joint publications.
The project described received funding from the National Research, Development,
and Innovation Office of Hungary (NKFIH, grants no. 152791 and K146096).
This paper was supported by the J\'anos Bolyai Research Scholarship
of the Hungarian Academy of Sciences.

\singlespacing
\bibliography{journals,new,master}

\end{document}